\documentclass[12pt]{article}
\usepackage{amsmath,amssymb,amsfonts, amsbsy, epsfig, subfig}
\usepackage[usenames]{color}
\usepackage{rotating}

\usepackage{boxedminipage}

\newcommand{\ignore}[1]{}{}

\newtheorem{theorem}{Theorem}[section]
\newtheorem{corollary}{Corollary}[section]
\newtheorem{proposition}{Proposition}[section]
\newtheorem{lemma}{Lemma}[section]
\newtheorem{example}{Example}[section]

\newtheorem{definition}{Definition}[section]

\newcommand{\beq}{\begin{equation}}
\newcommand{\eeq}{\end{equation}}
\newcommand{\beas}{\begin{eqnarray*}}
\newcommand{\eeas}{\end{eqnarray*}}
\newcommand{\bea}{\begin{eqnarray}}
\newcommand{\eea}{\end{eqnarray}}
\newcommand{\bei}{\begin{itemize}}
\newcommand{\eei}{\end{itemize}}
\newcommand{\ben}{\begin{enumerate}}
\newcommand{\een}{\end{enumerate}}
\newcommand{\bet}{\begin{theorem}}
\newcommand{\eet}{\end{theorem}}
\newcommand{\bel}{\begin{lemma}}
\newcommand{\eel}{\end{lemma}}
\newcommand{\bep}{\begin{proposition}}
\newcommand{\eep}{\end{proposition}}
\newcommand{\bed}{\begin{definition}}
\newcommand{\eed}{\end{definition}}
\newcommand{\bec}{\begin{corollary}}
\newcommand{\eec}{\end{corollary}}
\newcommand{\bex}{\begin{example}}
\newcommand{\eex}{\end{example}}
\newcommand{\qed}{\quad\hbox{\vrule width 4pt height 6pt depth 1.5pt}}
\newcommand{\RR}{I\!\! R}
\newcommand{\al}{\alpha}

\newcommand{\de}{\delta}
\newcommand{\ep}{\epsilon}

\newcommand{\argmin}{\mathop{\rm arg\min}}

\def\0{\boldsymbol{0}}

\def\a{\boldsymbol{a}}
\def\b{\boldsymbol{b}}

\def\R{\boldsymbol{R}}

\def\D{\boldsymbol{\textbf{D}}}
\def\A{\boldsymbol{A}}
\def\z{\boldsymbol{z}}
\def\W{\boldsymbol{W}}

\def\x{\boldsymbol{x}}
\def\be{\boldsymbol{\beta}}
\def\de{\boldsymbol{\delta}}
\def\X{\boldsymbol{X}}
\def\I{\boldsymbol{I}}

\def\va{\boldsymbol{\varepsilon}}
\def\S{\boldsymbol{\Sigma}}

\def\O{\boldsymbol{\Omega}}
\def\o{\boldsymbol{\omega}}
\def\m{\boldsymbol{\mu}}
\def\0{\boldsymbol{0}}

\def\al{\boldsymbol{\alpha}}
\def\x{\boldsymbol{x}}

\def\be{\boldsymbol{\beta}}
\def\de{\boldsymbol{\delta}}
\def\va{\boldsymbol{\varepsilon}}
\def\X{\boldsymbol{X}}
\def\x{\boldsymbol{x}}

\def\S{\boldsymbol{\Sigma}}
\def\O{\boldsymbol{\Omega}}
\def\m{\boldsymbol{\mu}}

\def\pr{\textsf{P}} 
\def\ep{\textsf{E}} 
\def\Cov{\textsf{Cov}} 
\def\Var{\textsf{Var}} 

\begin{document}

\title{Gaussian Graphical Model Estimation with False Discovery Rate Control\footnote{ Weidong Liu is Professor, Department of Mathematics, Institute of Natural Sciences and MOE-LSC, Shanghai Jiao Tong University, Shanghai, China (Email: liuweidong99@gmail.com).  The research was supported by NSFC, Grant No.11201298, the Program for Professor of Special Appointment (Eastern Scholar) at Shanghai Institutions of Higher Learning, Foundation for the Author of National Excellent Doctoral Dissertation
of PR China and  Program for New Century Excellent Talents in University. }
}
\author{Weidong Liu }

\date{}
\maketitle

\begin{abstract}
This paper studies the estimation of high dimensional Gaussian graphical model (GGM).
Typically,  the existing methods depend on regularization techniques. As a result, it is necessary to choose the regularized parameter.
However, the precise relationship between the regularized parameter and the number of false edges in  GGM estimation is unclear.  Hence, it is impossible to evaluate their performance rigorously. In this paper, we propose an alternative method  by a multiple testing procedure.  Based on our new test statistics  for conditional dependence, we propose a simultaneous testing procedure for conditional dependence in GGM. Our method can control
the false discovery rate (FDR) asymptotically.  The numerical performance of the proposed method  shows that our method works quite well.

\end{abstract}

\section{Introduction}

Estimation of dependency networks for high dimensional datasets is especially desirable in many scientific areas such as biology and sociology.
Gaussian graphical model (GGM) has proven to be a very powerful formalism to infer  dependence structures of various datasets.
GGM is an equivalent representation of  conditional dependence of jointly Gaussian random variables.
Inference on the structure of GGM is challenging when the dimension is greater than the sample size. Many classical methods do not work any more.

Let $\X=(X_{1},\ldots,X_{p})^{'}$ be a multivariate normal random vector with mean $\m$ and covariance matrix $\S$.   GGM is a graph $G=(V,E)$, where $V=\{X_{1},\ldots,X_{p}\}$ is the set of vertices and $E$ is the set of edges between vertices. There is an edge between $X_{i}$ and $X_{j}$ if and only if $X_{i}$ and $X_{j}$ are conditional dependent given $\{X_{k}, k\neq i,j\}$.
It is well-known  that estimating the structure of GGM is equivalent to recovering the support of precision matrix $\O=\S^{-1}$; see Lauritzen (1996).

The typical way on GGM  estimation depends on regularized optimizations. The past decade has witnessed significant developments on the
regularization method for various statistical problems. For example, in the context of variable selection,  Tibshirani (1996) introduced Lasso, which selects important variables in regression by solving
the least squares optimization with the $l_{1}$ regularization. Graphical-Lasso, an extension of Lasso to GGM estimation, was introduced by
 Yuan and Lin (2007), Friedman et al. (2008) and d'Aspremont
et al. (2008).  Graphical-Lasso estimates the support of precision matrix  by an $l_{1}$ penalized likelihood method. Theoretical properties of Graphical-Lasso can be found in Rothman et al. (2008) and Ravikumar et al. (2011). Other methods,  based on the $l_{1}$-minimization  technique, can be found in Meinshausen and Buhlmann (2006), Yuan (2010), Zhang (2010), Cai, et al. (2011), Liu, et al. (2012), Xue and Zou (2012). The nonconvex
penalties, such as SCAD function penalty (Fan
et al. (2009)), have also been considered in the context of GGM estimation.

It is well known that regularization  approaches often require the choice of tuning parameters. Large tuning parameters  often lead to sparse networks and they are powerless on finding  the  edges with small weights. On the other hand, small tuning parameters will generate many  false edges and  result in high false discovery rates.
The theory of the precise relationship between the number of false edges and the tuning parameter is very difficult to be derived.

A different way  on GGM estimation relies on  simultaneous tests
\begin{eqnarray}\label{test}
H_{0ij}:~\omega_{ij}=0\quad\mbox{versus\quad} H_{1ij}:~\omega_{ij}\neq 0
\end{eqnarray}
for $1\leq i<j\leq p$, where $\O=:(\omega_{ij})_{p\times p}$. An edge between $X_{i}$ and $X_{j}$ is included into the estimated network if and only if $H_{0ij}$ is rejected.
 When the dimension $p$ is fixed,
Drton and Perlman (2004)  proposed a multiple testing procedure to estimate GGM. They used the  Fisher's z transformations
of the sample partial correlation coefficients (SPCCs). A procedure on controlling the family-wise error was  developed. However, when the dimension $p$ is greater than the sample size, the sample partial correlation matrix is not even well defined. Hence, we do not have a natural
pivotal estimator as SPCCs so that the  asymptotic null distribution is easy to be derived. In high dimensional settings,
it becomes very challenging to estimate GGM by tests on the entries of precision matrix.

 In the present paper, we study the estimation of GGM by multiple tests (\ref{test}). We are particularly interested in high dimensional settings. The false discovery rate (FDR) is a useful measure  on evaluating the performance of  GGM estimation.  We will introduce a procedure called GGM estimation with FDR control (GFC).

 A basic step in hypothesis tests is the construction of test statistics.
 The sample partial correlation coefficients are not well defined when $p>n$.  Hence,  we introduce  new test statistics  suitable for high dimensional settings.  The new test statistics are based on a bias correction version of the sample covariance coefficients of residuals. They are shown to be asymptotically normal distributed under some sparsity conditions on $\O$. In addition to new test statistics, GFC carries out large-scale tests simultaneously. To this end, an adjustment for significance  levels is necessary. In this paper, we develop a multiple testing procedure with an adjustment for significance  levels and it controls the false discovery rate. The proposed procedure thresholds test statistics directly rather than p-values
 which were widely used (cf. Benjamini and Hochberg (1995)).
It is convenient for us to develop novel theoretical properties on FDR. We show that  GFC method controls both FDR and false discovery proportion (FDP) asymptotically.

In addition to its desirable theoretical properties, GFC method is computationally
very attractive for high dimensional data. The computational cost is the same as the neighborhood selection method by Meinshausen and Buhlmann (2006) or the CLIME method by Cai, et al. (2011). We only need to solve  $p$ regression equations with  Lasso or Dantzig selector. Numerical
performance of GFC is investigated by simulated data. Results
show that the procedure performs favorably in controlling  FDR and FDP.

The rest of the paper is organized as follows. In Section 2.2, we introduce new test statistics for conditional dependence.  GFC procedure is introduced in Section 2.3. In Section 3, we give limiting distributions of our test statistics. Theoretical results on GFC are also stated.
 Since GFC needs initial estimations of regression coefficients, we provide  their detailed  implementations in Section 4.
Numerical performance of the procedure is evaluated by
simulation studies in Section 5. The proofs of main results are delegated to Section 6.

\section{ Tests on conditional dependence}

We begin this section by introducing basic notations.
For any vector $\x$, Let $\x_{-i}$ denote $p-1$ dimensional vector by removing $x_{i}$ from $\x=(x_{1},\ldots,x_{p})^{'}$. For any $p\times q$ matrix $\A$, Let $\A_{i,-j}$ denote the $i$-th row of $\A$ with its $j$th entry being removed and $\A_{-i,j}$ denote
 the $j$-th column of $\A$ with its $i$th entry being removed. $\A_{-i,-j}$ denote a $(p-1)\times (q-1)$ matrix by removing the $i$-th row and $j$-th column of $\A$. Throughout,  define $|\x|_{0}=\sum_{j=1}^{p}I\{x_{j}\neq 0\}$,
$|\x|_{1}=\sum_{j=1}^{p}|x_{j}|$ and
$|x|_{2}=\sqrt{\sum_{j=1}^{p}x^{2}_{j}}$. For a matrix
$\A=(a_{ij})\in\RR^{p\times q}$, we define the element-wise $l_{\infty}$
norm $|\A|_{\infty}=\max_{1\leq i\leq p,1\leq j\leq q}|a_{ij}|$, the
spectral norm $\|\A\|_{2}=\sup_{|\textbf{x}|_{2}\leq
  1}|\A\textbf{x}|_{2}$ and the matrix $\ell_1$ norm $\|\A\|_{l_{1}}=\max_{1\leq
  j\leq q}\sum_{i=1}^{p}|a_{ij}|$. Let $\lambda_{\max}(\S)$ and $\lambda_{\min}(\S)$ denote the largest eigenvalue and the smallest eigenvalue of $\S$ respectively. $\I_{p}$ denotes a
$p\times p$ identity matrix. Let $\mathcal{H}_{0}=\{(i,j):\omega_{ij}=0,\quad 1\leq i<j\leq p\}$ and $\mathcal{H}_{1}=\{(i,j):\omega_{ij}\neq 0,\quad 1\leq i<j\leq p\}$.

 It is well known  that, for  $\X=(X_{1},\ldots,X_{p})^{'}\sim N(\m,\S)$,  we can write
\begin{eqnarray}\label{reg}
X_{i}=\alpha_{i}+\X_{-i}^{'}\be_{i}+\varepsilon_{i},
\end{eqnarray}
where $\varepsilon_{i}\sim N(0, \sigma_{ii}-\S_{i,-i}\S_{-i,-i}^{-1}\S_{-i,i})$ is independent of $\X_{-i}$, $\alpha_{i}=\mu_{i}-\S_{i,-i}\S_{-i,-i}^{-1}\m_{-i}$ and $(\sigma_{ij})_{p\times p}=\S$; see Anderson (2003).
The regression coefficients vector $\be_{i}$ and the error terms $\varepsilon_{i}$  satisfy
\begin{eqnarray*}
\be_{i}=-\omega_{ii}^{-1}\O_{-i,i}\quad\mbox{and\quad}\Cov(\varepsilon_{i},\varepsilon_{j})=\frac{\omega_{ij}}{\omega_{ii}\omega_{jj}}.
\end{eqnarray*}
We estimate GGM by recovering the support of $\S_{\varepsilon}$, the covariance matrix of $(\varepsilon_{1},\ldots,\varepsilon_{p})^{'}$.


\subsection{Test statistics for $H_{0ij}$}
In this subsection, we introduce new test statistics for $H_{0ij}$.
 Let $\textbf{X}=(\X_{1},\ldots,\X_{n})^{'}$, where $\X_{k}=(X_{k1},\ldots,X_{kp})^{'}$, $1\leq k\leq n$, be independent and identically distributed random samples from $\X$.
By (\ref{reg}), we can write
\begin{eqnarray*}
X_{ki}=\alpha_{i}+\textbf{X}_{k,-i}\be_{i}+\varepsilon_{ki},\quad 1\leq k\leq n,
\end{eqnarray*}
where $\textbf{X}_{k,-i}$ is  the $k$-th row of $\textbf{X}$ with its $i$th entry being removed and $\varepsilon_{ki}$ is independent with $\textbf{X}_{k,-i}$.
Let $\hat{\be}_{i}=(\hat{\beta}_{1,i},\ldots,\hat{\beta}_{p-1,i})^{'}$ be any estimators of $\be_{i}$ satisfying
\begin{eqnarray}\label{bt1}
\max_{1\leq i\leq p}|\hat{\be}_{i}-\be_{i}|_{1}=O_{\pr}(a_{n1})
\end{eqnarray}
and
\begin{eqnarray}\label{bt2}
\min\Big{\{}\lambda^{1/2}_{\max}(\S)\max_{1\leq i\leq p}|\hat{\be}_{i}-\be_{i}|_{2},\max_{1\leq i\leq p}\sqrt{(\hat{\be}_{i}-\be_{i})^{'}\hat{\S}_{-i,-i}(\hat{\be}_{i}-\be_{i})}\Big{\}}=O_{\pr}(a_{n2})
\end{eqnarray}
for some convergence rates $a_{n1}$ and $a_{n2}$, where $\hat{\S}=\frac{1}{n}\sum_{k=1}^{n}(\X_{k}-\bar{\X})(\X_{k}-\bar{\X})^{'}$ and $\bar{\X}=\frac{1}{n}\sum_{k=1}^{n}\X_{k}$.
Define the residuals by
\begin{eqnarray*}
\hat{\varepsilon}_{ki}=X_{ki}-\bar{X}_{i}-(\textbf{X}_{k,-i}-\bar{\textbf{X}}_{-i})\hat{\be}_{i}
\end{eqnarray*}
and the sample covariance coefficients between the residuals by
\begin{eqnarray}\label{a0}
\hat{r}_{ij}=\frac{1}{n}\sum_{k=1}^{n}\hat{\varepsilon}_{ki}\hat{\varepsilon}_{kj},
\end{eqnarray}
where $\bar{X}_{i}=\frac{1}{n}\sum_{k=1}^{n}X_{ki}$ and $\bar{\textbf{X}}_{-i}=\frac{1}{n}\sum_{k=1}^{n}\textbf{X}_{k,-i}$.
Our test statistics are based on a bias correction of $\hat{r}_{ij}$.
To this end, for $1\leq i<j\leq p$, define
\begin{eqnarray}\label{a2}
T_{ij}:=\frac{1}{n}\Big{(}\sum_{k=1}^{n}\hat{\varepsilon}_{ki}\hat{\varepsilon}_{kj}+\sum_{k=1}^{n}\hat{\varepsilon}_{ki}^{2}\hat{\beta}_{i,j}
+\sum_{k=1}^{n}\hat{\varepsilon}_{kj}^{2}\hat{\beta}_{j-1,i}\Big{)}.
\end{eqnarray}
It should be noted that the index is $j-1$ in $\hat{\beta}_{j-1,i}$ and $\hat{\be}_{i}$ is a $p-1$ dimensional vector.
Let
\begin{eqnarray}\label{b1}
b_{nij}=\omega_{ii}\hat{\sigma}_{ii,\varepsilon}+\omega_{jj}\hat{\sigma}_{jj,\varepsilon}-1,
 \end{eqnarray}
 where $(\hat{\sigma}_{ij,\varepsilon})_{1\leq i,j\leq p}=\frac{1}{n}\sum_{k=1}^{n}(\va_{k}-\bar{\va})(\va_{k}-\bar{\va})^{'}$, $\va_{k}=(\varepsilon_{k1},\ldots,\varepsilon_{kp})^{'}$ and $\bar{\va}=\frac{1}{n}\sum_{k=1}^{n}\va_{k}$. We will prove that
\begin{eqnarray*}
T_{ij}
=-b_{nij}\frac{\omega_{ij}}{\omega_{ii}\omega_{jj}}+\frac{\sum_{k=1}^{n}(\varepsilon_{ki}\varepsilon_{kj}-\ep \varepsilon_{ki}\varepsilon_{kj}) }{n}+O_{\pr}\Big{(}\lambda_{\max}(\S)a_{n2}^{2}+a_{n1}\sqrt{\frac{\log p}{n}}+\frac{\log p}{n}\Big{)}.
\end{eqnarray*}
And under
\begin{eqnarray}\label{c1}
 a_{n2}=o(n^{-1/4})\mbox{\quad and\quad } a_{n1}=o(1/\sqrt{\log p}),
\end{eqnarray}
 we will prove that
\begin{eqnarray}\label{a1}
\sqrt{\frac{n}{\hat{r}_{ii}\hat{r}_{jj}}}(T_{ij}+b_{nij}\frac{\omega_{ij}}{\omega_{ii}\omega_{jj}})\Rightarrow N\Big{(}0,1+\frac{\omega_{ij}^{2}}{\omega_{ii}\omega_{jj}}\Big{)}.
\end{eqnarray}
Note that, under $H_{0ij}$, the limiting distribution in (\ref{a1}) does not depend on any unknown parameter. Also, $b_{nij}\rightarrow 1$ in probability, uniformly in $1\leq i\leq j\leq p$.
Hence, for the  hypothesis test $H_{0ij}$, we shall use the following test statistic
\begin{eqnarray}\label{d1}
\hat{T}_{ij}=\sqrt{\frac{n}{\hat{r}_{ii}\hat{r}_{jj}}}T_{ij}.
\end{eqnarray}

The  estimators $\hat{\be}_{i}$, $1\leq i\leq p$, can be  Lasso estimators or  Dantizg selectors. Theoretical results on the convergence rates in (\ref{c1})
have been proved by many papers under various conditions. For example, for Dantizg selector, it can be proved by (46) and (47) that, under (C1) in Section 3,
(\ref{c1}) is satisfied when $\max_{1\leq i\leq p}|\be_{i}|_{0}=o\Big{(}\lambda_{\min}(\S)\frac{\sqrt{n}}{\log p}\Big{)}$. The same conclusion holds for the Lasso estimators. The detailed choices of $\hat{\be}_{i}$ will be given in Section 4.

\noindent{\bf Remark 1.} There are a number of recent papers  in
the  regression context where bias correction is used to derive p-values or confidence
intervals for the regression coefficients in the high-dimensional case; see Zhang and Zhang (2011), B\"{u}hlmann (2012), van de Geer, Bhlmann and Ritov (2013),
Javanmard and Montanari (2013). When applying their methods in GGM estimation, we briefly discuss the difference between our method and theirs.
For every $i$, to get the p-values for the components of $\be_{i}$, their methods need to estimate the $(p-1)\times (p-1)$ precision matrix of $\X_{-i}$.
So, to derive the p-values for all of the components of $\be_{i}$, $1\leq i\leq p$, their methods need to estimate $p$  precision matrices with dimension $(p-1)\times (p-1)$. This requires a huge computational cost. Our method only needs the initial estimators for $\be_{i}$. No additional precision matrix estimator is required.

\subsection{GGM estimation with FDR control}

With the new test statistic $\hat{T}_{ij}$,  we can carry out $(p^{2}-p)/2$ tests (\ref{test}) simultaneously and control FDR as follow. Let $t$ be the threshold level such that $H_{0ij}$ is rejected if $|\hat{T}_{ij}|\geq t$.
The false discovery rate and false discovery proportion are defined by
\begin{eqnarray*}
\text{FDP}(t)=\frac{\sum_{(i,j)\in\mathcal{H}_{0}}I\{|\hat{T}_{ij}|\geq t\}}{\max\{\sum_{1\leq j<j\leq p}I\{|\hat{T}_{ij}|\geq t\},1\}},\quad \text{FDR}(t)=\ep [\text{FDP}(t)].
\end{eqnarray*}
 A "good" threshold level $t$  makes as many as true alternative hypothesis be rejected and remains the FDR/FDP be controlled at a pre-specified level $0<\alpha<1$. So an ideal choice of $t$ is

\begin{eqnarray*}
\hat{t}_{o}=\inf\{0\leq t\leq 2\sqrt{\log p}: \frac{\sum_{(i,j)\in\mathcal{H}_{0}}I\{|\hat{T}_{ij}|\geq t\}}{\max\{\sum_{1\leq j<j\leq p}I\{|\hat{T}_{ij}|\geq t\},1\}}\leq \alpha\},
\end{eqnarray*}
where $\mathcal{H}_{0}=\{(i,j):\omega_{ij}=0,\quad 1\leq i<j\leq p\}$. In the definition of $\hat{t}_{o}$, $t$ is restricted to $[0,2\sqrt{\log p}]$ because $\pr(\max_{(i,j)\in\mathcal{H}_{0}}|\hat{T}_{ij}|\geq 2\sqrt{\log p})\rightarrow 0$ by the proof in Section 6.
Since $\mathcal{H}_{0}$ is unknown, we shall use an estimator of $\sum_{(i,j)\in\mathcal{H}_{0}}I\{|\hat{T}_{ij}|\geq t\}$. As we will prove in Section 6, an  accurate approximation  for $\sum_{(i,j)\in\mathcal{H}_{0}}I\{|\hat{T}_{ij}|\geq t\}$ is  $2(1-\Phi(t))|\mathcal{H}_{0}|$, where $\Phi(t)=\pr(N(0,1)\leq t)$. Moreover, $\mathcal{H}_{0}$ can be estimated by $(p^{2}-p)/2$ due to the sparsity of $\O$. This leads to the following procedure.

\begin{center}
\begin{boxedminipage}{1.0\textwidth}
{\bf GFC procedure.}  Calculate test statistics $\hat{T}_{ij}$ in (\ref{d1}). Let $0<\alpha<1$ and
\begin{eqnarray}\label{a3}
\hat{t}=\inf\Big{\{}0\leq t\leq 2\sqrt{\log p}: \frac{G(t)(p^{2}-p)/2}{\max\{\sum_{1\leq i<j\leq p}I\{|\hat{T}_{ij}|\geq t\},1\}}\leq \alpha\Big{\}},
\end{eqnarray}
where $G(t)=2-2\Phi(t)$. If $\hat{t}$ in (\ref{a3}) does not exist, then let $\hat{t}=2\sqrt{\log p}$. For $1\leq i<j\leq p$, we reject $H_{0ij}$ if $|\hat{T}_{ij}|\geq \hat{t}$.
\end{boxedminipage}
\end{center}\vspace{5mm}

In GFC procedure, the estimators $\hat{\be}_{i}$, $1\leq i\leq p$, are needed. As mentioned earlier, we can use the Lasso estimators or the Dantizg selectors. Both of them require the choice of tuning parameters.  In Section 4, we will propose a  method on the choice of tuning parameters, which is particularly suitable for our multiple testing problem.

For general multiple testing problems, Liu and Shao (2012) developed a  procedure that controls the false discovery rate. They proposed to threshold
test statistics directly rather than the true p-values as in Benjamini and Hochberg (1995), because
 the true p-values are unknown in practice. Additionally, to control FDR, the Benjamini-Hochberg  method requires the independence or some kind of positive regression dependency between  p-values. Our test statistics do not meet such conditions. By thresholding the test statistics directly as in Liu and Shao (2012), we shall show that $\text{FDR}(\hat{t})\rightarrow \alpha$ and $\text{FDP}(\hat{t})\rightarrow\alpha$ in probability. It should be pointed out that
  Liu and Shao (2012) imposed the dependence condition among the test statistics. In GGM estimation, it is more natural to impose the dependence condition on
  the precision matrix. To this end, we need many novel techniques in the proof.

\section{Theoretical results}

In this section, we  will show that  GFC procedure can control
the false discovery rate asymptotically at any pre-specified level.  \\


{\bf (C1).} Let $\X\sim N(\m,\S)$. Suppose that $\max_{1\leq i\leq p}\sigma_{ii}\leq c_{0}$ and $\max_{1\leq i\leq p}\omega_{ii}\leq c_{0}$ for some constant $c_{0}>0$. Assume that
 $\log p=o(n)$.\vspace{4mm}

  Since $\sigma_{ii}\omega_{ii}\geq 1$, (C1)
 implies that $\min_{1\leq i\leq p}\omega_{ii}\geq c^{-1}_{0}$ and $\min_{1\leq i\leq p}\sigma_{ii}\geq c^{-1}_{0}$.
We give the asymptotic distribution of  $\hat{T}_{ij}$, which is useful in testing
 a single $H_{0ij}:$ $\omega_{ij}=0$.

 \begin{proposition}
 Suppose that (C1) holds. Let $\hat{\be}_{i}$ be any estimator satisfying (\ref{bt1}), (\ref{bt2}) and (\ref{c1}).  Then, we  have
 \begin{eqnarray*}
 \sqrt{\frac{n}{\hat{r}_{ii}\hat{r}_{jj}}}(T_{ij}+b_{nij}\frac{\omega_{ij}}{\omega_{ii}\omega_{jj}})\Rightarrow N(0,1+\frac{\omega_{ij}^{2}}{\omega_{ii}\omega_{jj}})
 \end{eqnarray*}
 as $(n,p)\rightarrow\infty$,
 where the convergence in distribution is uniformly in $1\leq i<j\leq p$.
 \end{proposition}

Let the false discovery proportion and false discovery rate of GFC be defined by
\begin{eqnarray*}
\mbox{FDP}=\frac{\sum_{(i,j)\in\mathcal{H}_{0}}I\{|\hat{T}_{ij}|\geq \hat{t}\}}{\max ( \sum_{1\leq i<j\leq p}I\{|\hat{T}_{ij}|\geq \hat{t}\},1) },\quad\mbox{FDR} = \ep (\mbox{FDP}).
\end{eqnarray*}
Recall that $\mathcal{H}_{0}=\{(i,j):\omega_{ij}=0,\quad 1\leq i<j\leq p\}$. Let $q_{0}=Card(\mathcal{H}_{0})$ be the cardinality of $\mathcal{H}_{0}$ and $q=(p^{2}-p)/2$. For a constant $\gamma>0$ and $1\leq i\leq p$, define $$\mathcal{A}_{i}(\gamma)=\{j: 1\leq j\leq p,~j\neq i,~|\omega_{ij}|\geq (\log p)^{-2-\gamma}\}.$$ Theorem \ref{th1} shows that GFC controls FDP and FDR at level $\alpha$ asymptotically.

\begin{theorem}\label{th1} Let $p\leq n^{r}$ for some $r>0$. Suppose that for some $\delta>0$,
\begin{eqnarray}\label{a5}
Card\Big{\{}(i,j): 1\leq i< j\leq p,~~\frac{|\omega_{ij}|}{\sqrt{\omega_{ii}\omega_{jj}}}\geq 4\sqrt{\log p/n}\Big{\}}\geq \Big{(}\frac{1}{\sqrt{8\pi}\alpha}+\delta\Big{)}\sqrt{\log_{2}p}.
\end{eqnarray}
Assume that $q_{0}\geq cp^{2}$ for some $c>0$ and $\hat{\be}_{i}$ satisfies (\ref{bt1}), (\ref{bt2}) and
\begin{eqnarray}\label{cd2}
a_{n1}=o(1/\log p) \mbox{\quad and\quad } a_{n2}=o((n\log p)^{-1/4}).
\end{eqnarray}
Under (C1) and $\max_{1\leq i\leq p}Card(\mathcal{A}_{i}(\gamma))=O(p^{\rho})$ for some $\rho<1/2$ and $\gamma>0$, we have
\begin{eqnarray*}
\lim_{(n,p)\rightarrow\infty}\frac{\text{FDR}}{\alpha q_{0}/q}=1\mbox{\quad and\quad}  \frac{\text{FDP}}{\alpha q_{0}/q}\rightarrow 1\mbox{~~in probability}
\end{eqnarray*}
as $(n,p)\rightarrow\infty$.
\end{theorem}

The dimension $p$ can be much larger than the sample size because $r$ can be arbitrarily large. Note that $q_{0}\geq cp^{2}$ is a natural condition. If $q_{0}=o(p^{2})$, then almost all of
$\omega_{ij}$ are nonzero. Hence, rejecting all the hypothesis tests leads to $\text{FDR}\rightarrow 0$. The condition $\max_{1\leq i\leq p}Card(\mathcal{A}_{i}(\gamma))=O(p^{\rho})$  is also mild. For example, if $p\geq n^{\delta}$ for some  $\delta>1$ and $\O$ is a $s_{n,p}$-sparse matrix with $s_{n,p}=O(\sqrt{n})$ (i.e. the number of nonzero entries in each row is no more than $s_{n,p}$), then this condition holds. The sparsity $s_{n,p}=O(\sqrt{n})$ is often imposed in the literature on precision matrix estimation.

The technical condition (\ref{a5}) is  used to ensure  $\Var(\sum_{(i,j)\in\mathcal{H}_{0}}I\{|\hat{T}_{ij}|\geq \hat{t}\})\rightarrow \infty$
which is almost necessary for
\begin{eqnarray}\label{c2}
\sum_{(i,j)\in\mathcal{H}_{0}}I\{|\hat{T}_{ij}|\geq \hat{t}\}/(|\mathcal{H}_{0}|G(\hat{t}))\rightarrow 1
 \end{eqnarray}
 in probability. We believe (\ref{c2})
 is nearly necessary for the false discovery proportion $\frac{\text{FDP}}{\alpha q_{0}/q}\rightarrow 1$ in probability. On the other hand, the condition for controlling FDR may be weaker than that for controlling FDP. Even if (\ref{c2}) is violated,
 the false discovery rate may still be controlled at level $\alpha$. Hence, it is possible that (\ref{a5}) is not needed for  FDR results. In addition,
 (\ref{a5}) is not strong because the total number of hypothesis tests is $(p^{2}-p)/2$ and we only require a few standardized off-diagonal entries of $\O$ have magnitudes exceeding $4\sqrt{\log p/n}$.



\section{Data-driven choice of $\hat{\be}_{i}$}

GFC requires to choose the estimators of $\be_{i}$. There are lots of literature on the estimation of high dimensional regression coefficients. In this paper, we use the popular  Dantizg selector and Lasso estimator. Some other recent procedures such as scaled-Lasso (Sun and Zhang, 2012) and Square-root Lasso (Belloni, Chernozhukov and Wang, 2011) can also be used and similar theoretical results as Proposition \ref{pro2} and \ref{pro3-3} can be established.

{\bf Dantizg selector for $\hat{\be}_{i}$.} Dantizg selector estimates $\hat{\be}_{i}$ by solving the following optimization problems
\begin{eqnarray}\label{so1}
\hat{\be}_{i}(\delta)=\argmin\{|\o|_{1}\quad\mbox{subject to\quad}|\D_{i}^{-1/2}\hat{\S}_{-i,-i}\o-\D_{i}^{-1/2}\hat{\a}|_{\infty}\leq \lambda_{ni1}(\delta)\}
\end{eqnarray}
for $1\leq i\leq p$,
where $\D_{i}=diag(\hat{\S}_{-i,-i})$, $\hat{\a}=\frac{1}{n}\sum_{k=1}^{n}(\textbf{X}_{k,-i}-\textbf{X}_{-i})^{'}(X_{k,i}-\bar{X}_{i})$ and
\begin{eqnarray*}
\lambda_{ni1}(\delta)=\delta\sqrt{\frac{\hat{\sigma}_{ii,X}\log p}{n}}
 \end{eqnarray*}
for $\delta>0$, where $\hat{\sigma}_{ii,X}=\frac{1}{n}\sum_{k=1}^{n}(X_{ki}-\bar{X}_{i})^{2}$.
We can let $\delta=2$ which is fully specified and has theoretical interest. For finite sample sizes, we will propose a more useful data-driven choice for $\delta$  in (\ref{ats}).

 \begin{proposition}\label{pro2}  Suppose that (C1) holds and $\max_{1\leq i\leq p}|\be_{i}|_{0}=o\Big{(}\lambda_{\min}(\S)\frac{\sqrt{n}}{(\log p)^{3/2}}\Big{)}$. For $\delta=2$ in (\ref{so1}), we have $\hat{\be}_{i}(2)$, $1\leq i\leq p$, satisfy (\ref{bt1}), (\ref{bt2}) and (\ref{cd2}).
 \end{proposition}

 {\bf  Lasso estimator for $\hat{\be}_{i}$.} The coefficients $\hat{\be}_{i}$ can be estimated by Lasso as follow:
\begin{eqnarray}\label{so2-2}
\hat{\be}_{i}(\delta)=\D_{i}^{-1/2}\hat{\al}_{i}(\delta),
\end{eqnarray}
where
\begin{eqnarray*}
\hat{\al}_{i}(\delta)=\argmin_{\al\in \textbf{R}^{p-1}} \Big{\{} \frac{1}{2n}\sum_{k=1}^{n}(X_{ki}-\bar{X}_{i}-(\textbf{X}_{k,-i}-\bar{\textbf{X}}_{-i})\D_{i}^{-1/2}\al)^{2}+\lambda_{ni1}(\delta)|\al|_{1}\Big{\}}.
\end{eqnarray*}
The following proposition shows that for any $\delta>2$, (\ref{cd2}) is satisfied.  The data-driven choice for $\delta$  is given in (\ref{ats}).

 \begin{proposition}\label{pro3-3} Suppose that (C1) holds and $\max_{1\leq i\leq p}|\be_{i}|_{0}=o\Big{(}\lambda_{\min}(\S)\frac{\sqrt{n}}{(\log p)^{3/2}}\Big{)}$. For any $\delta>2$ in (\ref{so2-2}), we have $\hat{\be}_{i}(\delta)$, $1\leq i\leq p$, satisfy  (\ref{bt1}), (\ref{bt2}) and (\ref{cd2}).
 \end{proposition}

{\bf Data-driven choice of $\delta$.} As in many regularization approaches, the choice $\delta\geq 2$ is often large. Hence, in this paper, we propose to select $\delta$  adaptively by  data.  We let $\hat{\be}_{i}(\delta)$ be the solution to (\ref{so1}) or (\ref{so2-2}) and then obtain the statistics $\hat{T}_{ij}(\delta)$, $1\leq i<j\leq p$. As noted in Section 2.3,  GFC works because for good estimators $\hat{\be}_{i}(\delta)$, $1\leq i\leq p$, $\sum_{(i,j)\in\mathcal{H}_{0}}I\{|\hat{T}_{ij}(\delta)|\geq t\}$ will be close to $|\mathcal{H}_{0}| G(t)$. Hence, an oracle choice of $\delta$ can be
 \begin{eqnarray}\label{a10}
 \hat{\delta}_{o}=\argmin_{0\leq \delta\leq 2}\int_{\tau_{p}}^{1}\Big{(}\frac{\sum_{(i,j)\in\mathcal{H}_{0}}I\{|\hat{T}_{ij}(\delta)|\geq \Phi^{-1}(1-\frac{\alpha}{2})\}}{\alpha|\mathcal{H}_{0}|}-1\Big{)}^{2}d\alpha,
 \end{eqnarray}
where $\tau_{p}=G(2\sqrt{\log p})$. $\mathcal{H}_{0}$ is unknown, however. Since $\O$ is sparse,  $|\mathcal{H}_{0}|$ is close to $(p^{2}-p)/2$.
So a good choice of $\delta$ should minimize the following error
\begin{eqnarray}\label{a11}
\int_{\tau}^{1}\Big{(}\frac{\sum_{1\leq i\neq j\leq p}I\{|\hat{T}_{ij}(\delta)|\geq \Phi^{-1}(1-\frac{\alpha}{2})\}}{\alpha(p^{2}-p)}-1\Big{)}^{2}d\alpha,
 \end{eqnarray}
where $\tau>0$ is a fixed number bounded away from zero. The constraint $\alpha\geq \tau$ aims to ensure the nonzero entries part $\sum_{(i,j)\in \mathcal{H}_{1}}I\{|\hat{T}_{ij}(\delta)|\geq \Phi^{-1}(1-\frac{\alpha}{2})\}=o(\alpha (p^{2}-p))$. In our choice, we let  $\tau=0.3$. This leads to the final choice of $\delta$ by discretizing the integral as follow:
\begin{eqnarray}\label{ats}
\hat{\delta}=\hat{j}/N,\quad\hat{j}=\argmin_{0\leq j\leq 2N}\sum_{k=3}^{9}\Big{(}\frac{\sum_{1\leq i\neq j\leq p}I\{|\hat{T}_{ij}(j/N)|\geq \Phi^{-1}(1-\frac{k}{20})\}}{k(p^{2}-p)/10}-1\Big{)}^{2},
\end{eqnarray}
where $N$ is an integer number that can be pre-specified.
 Finally, we use $\hat{\be}_{i}(\hat{\de})$ as the estimator of $\be_{i}$.
 Deriving theoretical properties for $\hat{\delta}$ is important. We
 leave this as a future work.

\section{Numerical results}

In this section, we carry out simulations to examine the performance of GFC by the following graphs.

\begin{itemize}
\item{{\bf Band graph}.  $\O=(\omega_{ij})$, where $\omega_{i,i+1}=\omega_{i+1,i}=0.6$, $\omega_{i,i+2}=\omega_{i+2,i}=0.3$, $\omega_{ij}=0$ for $|i-j|\geq 3$. $\O$ is a $5$-sparse matrix. }

\item{ {\bf Hub graph}. There are $p/10$ rows with sparsity $11$. The rest every row has sparsity $2$. To this end, we let $\O_{1}=(\omega_{ij})$,
$\omega_{ij}=\omega_{ji}=0.5$ for $i=10(k-1)+1$ and $10(k-1)+2\leq j\leq 10(k-1)+10$, $1\leq k\leq p/10$. The diagonal $\omega_{ii}=1$ and others entries are zero. Finally, we let $\O=\O_{1}+(|\min(\lambda_{\min})|+0.05)\I_{p}$ to make the matrix be positive definite.}

\item{ {\bf Erd\"{o}s-R\'{e}nyi random graph}. There is an edge between each pair of nodes with probability
$\min(0.05,5/p)$ independently. Let $\omega_{ij}=u_{ij}*\delta_{ij}$, where $u_{ij}\sim U(0.4,0.8)$ is the uniform random variable and $\delta_{ij}$ is the
Bernoulli random variable with success probability 0.05. $u_{ij}$ and $\delta_{ij}$ are independent. Finally, we let $\O=\O_{1}+(|\min(\lambda_{\min})|+0.05)\I_{p}$ such that the matrix is positive definite.}


\end{itemize}

For each model, we generate $n=100$ random samples with $\X_{k}\sim N(\m,\S)$, $\S=\O^{-1}$ and $p=50,100,200, 400$. We use the Dantizg selector and Lasso to estimate $\be_{i}$  in GFC and denote the corresponding procedures by GFC-Dantizg and GFC-Lasso. The tuning parameter $\lambda_{ni1}(\hat{\delta})$ is  given in Section 4 with $N=20$. The simulation results are based on 100 replications.  As we can see from Table 1, the FDRs of GFC-Dantizg for Band graph and Erd\"{o}s-R\'{e}nyi (E-R) random graph are close to $\alpha$. The FDRs  for Hub graph are somewhat smaller than $\alpha$. For all three graphs, the  FDRs can be effectively controlled below the level $\alpha$. Similarly, GFC-Lasso can control FDR at the level $\alpha$.
The FDPs of  GFC-Dantizg in 100 replications are plotted in Figure 1 with $p=200$. For the reason of space, we give the other figures for $p=50,100,400$ and
  GFC-Lasso in the supplemental material Liu (2013). We can see from these figures that most of FDPs are concentrated around the  FDRs.

In Figure 2, we plot the  FDPs for all GFC-Dantizg estimators with $p=200$, $\alpha=0.2$ and $\hat{\be}_{i}(j/20)$, $1\leq j\leq 40$. The histograms of $\hat{j}$ are plotted in Figure 3. We use $\widehat{\text{FDR}}(j)$ to denote the false discovery rates for GFC-Dantizg with $\hat{\be}(j/20)$. As we can see from Figure 2, there always exist several $j$ such that $\widehat{\text{FDR}}(j)$ are well controlled at level $\alpha=0.2$. From the histograms of $\hat{j}$ in Figure 3, we see that $\hat{j}$ in Section 4 can always take the values of these $j$'s for all three graphs. Similar phenomenon can be observed in GFC-Lasso; see the supplemental material Liu (2013).


We examine the power of GFC on controlling FDR.
Based on 100 replications, the average  powers are defined by
\begin{eqnarray*}
\text{Average}\Big{\{}\frac{\sum_{(i.j)\in\mathcal{H}_{1}}I\{|\hat{T}_{ij}|\geq \hat{t}\}}{Card(\mathcal{H}_{1})}\Big{\}}.
\end{eqnarray*}
We state the numerical results in Table 2. The power increases when $\alpha$ increases. For the Hub graph, the powers are close to one. For the Band graph, GFC-Dantizg can also effectively detect the  edges and GFC-Lasso is more powerful than GFC-Dantizg. For the
 Erd\"{o}s-R\'{e}nyi random graph, GFC has non-trivial powers when $p=50$, $100$ and $200$. The powers are low when $p=400$. This mainly dues to the very small magnitude of $\omega_{ij}$. Actually, all of $\frac{\omega_{ij}}{\sqrt{\omega_{ii}\omega_{jj}}}$ belong to the interval $(0.1275,0.255)$ when $p=400$. So it is very difficult to detect such small nonzero entries.

 Finally, we compare GFC with the Graphical Lasso (Glasso) which estimates the graph by solving the following optimization problem:
 \begin{eqnarray*}
\hat{\O}(\lambda_{n}):=\argmin_{\O\succ
    0}\{\langle\O,\hat{\S}_{n}\rangle-\log\det(\O)+\lambda_{n}\|\O\|_{1}\}.
\end{eqnarray*}
As in Rothman, et al. (2008), Fan, Feng and Wu (2009) and Cai, Liu and Luo (2011),
the tuning parameter $\lambda_{n}$ is selected by the popular cross validation method. To this end,
we generate another $n=100$ training samples from $\X$ and let $\hat{\S}_{train}$ be the sample covariance matrix from the training samples.
We choose the following the tuning parameter
\begin{eqnarray*}
\lambda_{n}=\hat{k}/50,\quad \hat{k}=\argmin_{1\leq k\leq 200} \{\langle\hat{\O}(k/50),\hat{\S}_{train}\rangle-\log\det(\hat{\O}(k/50)).
\end{eqnarray*}
The empirical false discovery rates and the standard deviations are stated in Table 3. We can see that for all three graphs the FDRs of Glasso are quite close to 1.
This indicates that Glasso with the cross validation method fails to control the false discovery rate. We next examine the power of Glasso.
Since the power of Glasso depends on the choice of $\lambda_{n}$, we plot all of the FDRs and the average powers for $\hat{\O}(\lambda_{n})$ with $\lambda_{n}=\frac{1}{50},\frac{2}{50},\ldots,\frac{200}{50}$ in Figure 4 with $p=200$. Other figures for $p=50,100,400$ are given in the supplemental material Liu (2013).  As we can see from these figures, for the Band graph and ER graph,
the powers are quite low ($\leq 0.05$) if the FDRs $\leq 0.2$. Hence, for these two graphs, GFC significantly outperforms Glasso even we know the
oracle choice of the tuning parameter for Glasso. It is also interesting to see that, for the Hub graph, the power of Glasso is close to one even
the FDRs are small. This phenomenon is similar to that of GFC which also performs quite well  for the Hub graph.

\begin{table}[hptb]\tiny\addtolength{\tabcolsep}{-4pt}
  \begin{center}
    \caption{{\bf Empirical false discovery rates}}
    \begin{tabular}{|c | @{\hspace{2em}}c@{\hspace{1em}}c@{\hspace{1em}}c@{\hspace{1em}}c|@{\hspace{1em}}c@{\hspace{1em}} c@{\hspace{1em}} c@{\hspace{1em}} c| }
      \hline
      & \multicolumn{4}{c}{$\alpha=0.1$}&\multicolumn{4}{c|}{$\alpha=0.2$} \\
      \hline
    $p$  &  50 &100& 200& 400& 50&
      100 & 200 & 400\\
           \hline
       & \multicolumn{6}{c}{GFC-Dantizg}&\multicolumn{2}{c|}{} \\
      \hline
      Band     &   0.0899&   0.1085  &  0.1160    &   0.1168                        &0.1738    &0.1991     &     0.2103  & 0.2035 \\
Hub    &  0.0722    &  0.0599        & 0.0557       &   0.0459                          &0.1651    &0.1415   & 0.1369  &0.1154\\
E-R   &  0.1174 &  0.0887  & 0.0747                &  0.0892                           &0.2099    &0.1738     &  0.1516  &  0.1703 \\
      \hline
      & \multicolumn{6}{c}{GFC-Lasso}&\multicolumn{2}{c|}{} \\
      \hline
      Band     &   0.0849&   0.0768  &  0.0801         &    0.0842                 &0.1759    &0.1650     &    0.1707   &0.1718\\
Hub    &  0.0917    & 0.0835        & 0.0766          &   0.0708                       &0.1937    &0.1852    & 0.1693   &0.1560\\
E-R   &  0.1038 &  0.0967   & 0.1011          &       0.1180                            &0.2149     &0.1963      & 0.2083  &  0.2297  \\

      \hline
    \end{tabular}

    \label{tb:simu3}
  \end{center}
\end{table}

\begin{table}[hptb]\tiny\addtolength{\tabcolsep}{-4pt}
  \begin{center}
    \caption{{\bf Power of GFC (SD)}}
    \begin{tabular}{|c | @{\hspace{2em}}c@{\hspace{1em}}c@{\hspace{1em}}c@{\hspace{1em}}c|@{\hspace{1em}}c@{\hspace{1em}} c@{\hspace{1em}} c@{\hspace{1em}} c| }
      \hline
      & \multicolumn{4}{c}{$\alpha=0.1$}&\multicolumn{4}{c|}{$\alpha=0.2$} \\
      \hline
    $p$  &  50 &100& 200& 400& 50&
      100 & 200 &400\\
           \hline
       & \multicolumn{6}{c}{\qquad\qquad\qquad  GFC-Dantizg}&\multicolumn{2}{c|}{} \\
      \hline
      Band    &   0.7934(0.0447)&   0.7182(0.0368)  &  0.6688(0.0255)    &     0.6265(0.0151)                    &0.8547(0.0430)    &0.7937(0.0409)  & 0.7399(0.0283)  & 0.6865(0.0157) \\
Hub     &  0.9607(0.0503)    &  0.9767(0.0208)        & 0.9776(0.0140)    &     0.9778(0.0087)                    &0.9767(0.0384)    &0.9877(0.0139)   & 0.9873(0.0096) & 0.9868(0.0074)\\
E-R   &  0.7319(0.0652) &  0.3596(0.0445)  & 0.2623(0.0249)      &     0.1416(0.0140)                       &0.7943(0.0551)    &0.4693(0.0448)     &  0.3505(0.0240) & 0.2051(0.0177)  \\
      \hline
      & \multicolumn{6}{c}{\qquad\qquad\qquad GFC-Lasso}&\multicolumn{2}{c|}{} \\
      \hline
      Band     &   0.8814(0.0365)&   0.8489(0.0244)  &  0.8027(0.0215)  & 0.7491(0.0149)                         &0.9227(0.0306)    &0.8939(0.0234)  & 0.8490(0.0172) & 0.7955(0.0155) \\
Hub    &  0.9224(0.0647)    &  0.9202(0.0389)         &0.9202(0.0323)     &   0.9327(0.0181)                    &0.9553(0.0456)    &0.9531(0.0308)   & 0.9513(0.0218) &  0.9570(0.0132)\\
E-R  &  0.7629(0.0561) &  0.4178(0.0429)   & 0.3014(0.0266)        &    0.1596(0.0149)                             &0.8265(0.0550)     &0.5294(0.0412)      &  0.4063(0.0258)  &  0.2390(0.0168)  \\

      \hline
    \end{tabular}

    \label{tb:simu3}
  \end{center}
\end{table}

\begin{table}[hptb]\footnotesize\addtolength{\tabcolsep}{-4pt}
  \begin{center}
    \caption{{\bf Empirical false discovery rates (SD) for Glasso}}
    \begin{tabular}{|c | @{\hspace{2em}}c@{\hspace{1em}}c@{\hspace{1em}}c@{\hspace{1em}}c|}
      \hline

    $p$  &  50 &100& 200& 400\\
      \hline
      Band    &   0.8449(0.0073)&   0.8887(0.0035)  &  0.9156(0.0022)  & 0.9354(0.0020)  \\
Hub    &  0.8622(0.0101)    &  0.9074(0.0055)        & 0.9333(0.0013)  & 0.9509(0.0010)  \\
E-R  &  0.8513(0.0154) &  0.8257(0.0042)  & 0.8564(0.0253)   &0.8692(0.0024) \\
\hline
    \end{tabular}

    \label{tb:simu3}
  \end{center}
\end{table}

\begin{figure}[htbp]
\centering
 \subfloat[][Band graph]{\includegraphics[width=0.3\textwidth]{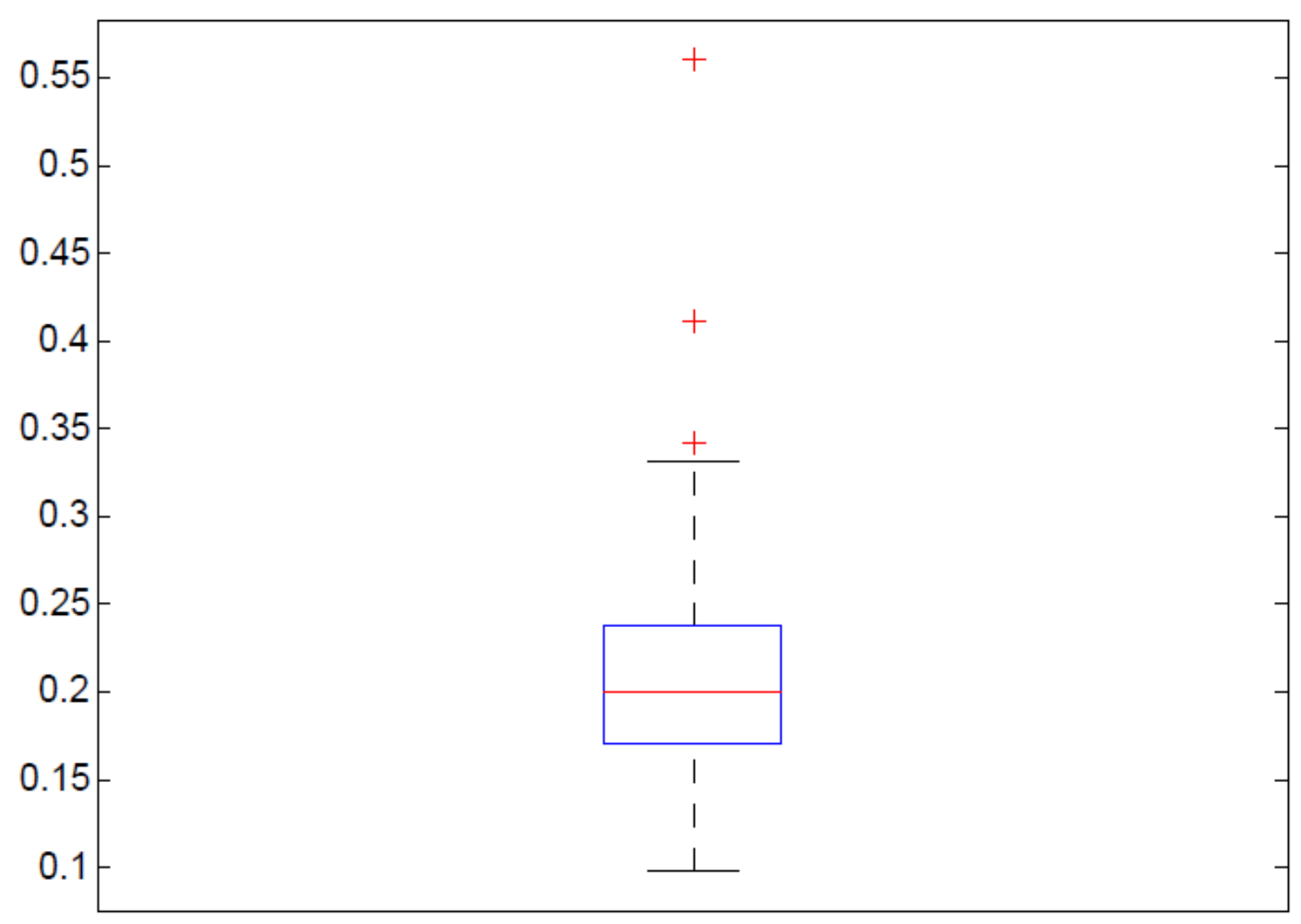} }
  \subfloat[][Hub graph]{\includegraphics[width=0.3\textwidth]{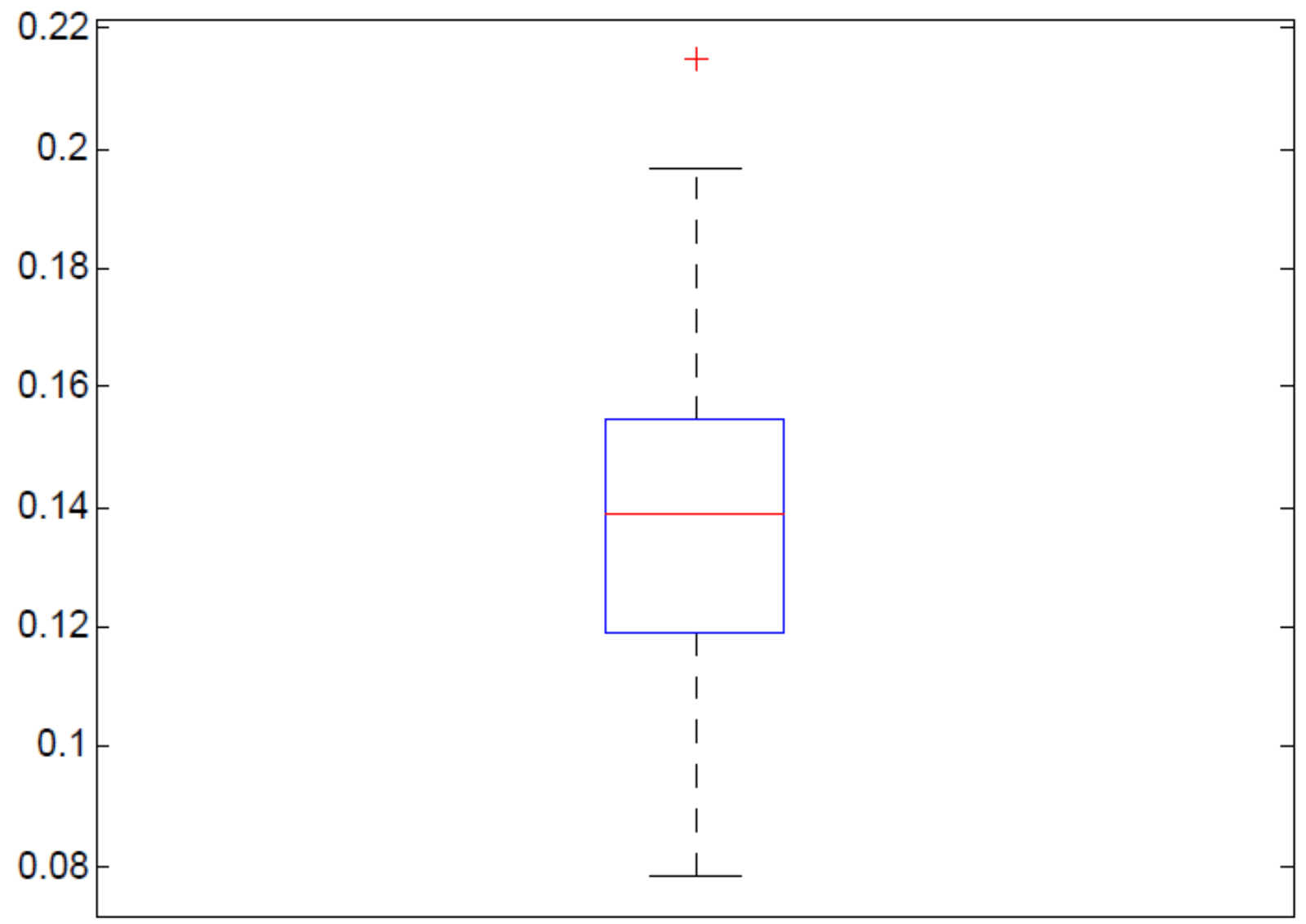} }
    \subfloat[][E-R graph]{\includegraphics[width=0.3\textwidth]{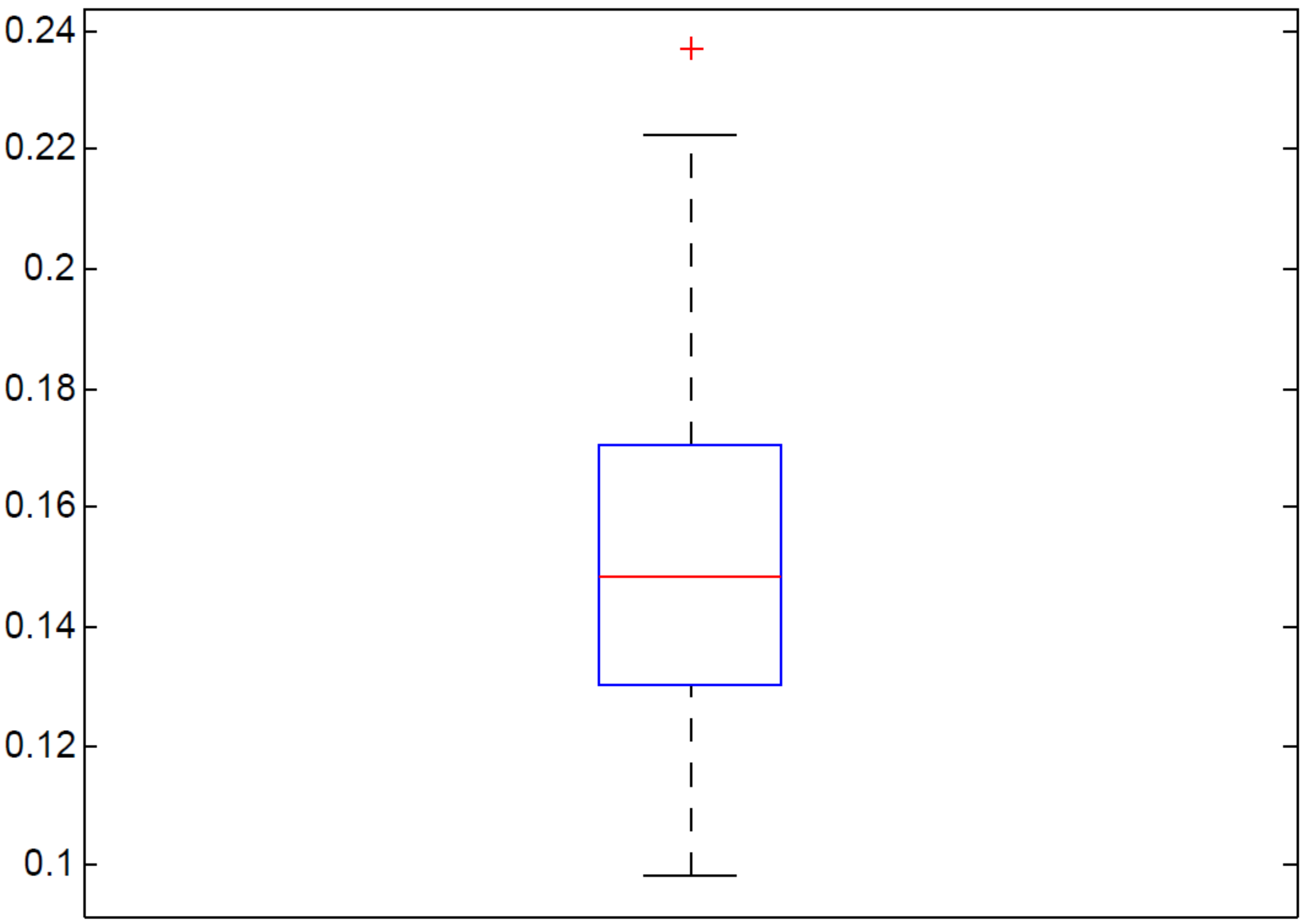} }
 \caption{FDP (GFC-Dantizg, $p=200$ and $\alpha=0.2$)}
 \label{fig:p50}
\end{figure}

\begin{figure}[htbp]
\centering
 \subfloat[][Band graph]{\includegraphics[width=0.3\textwidth]{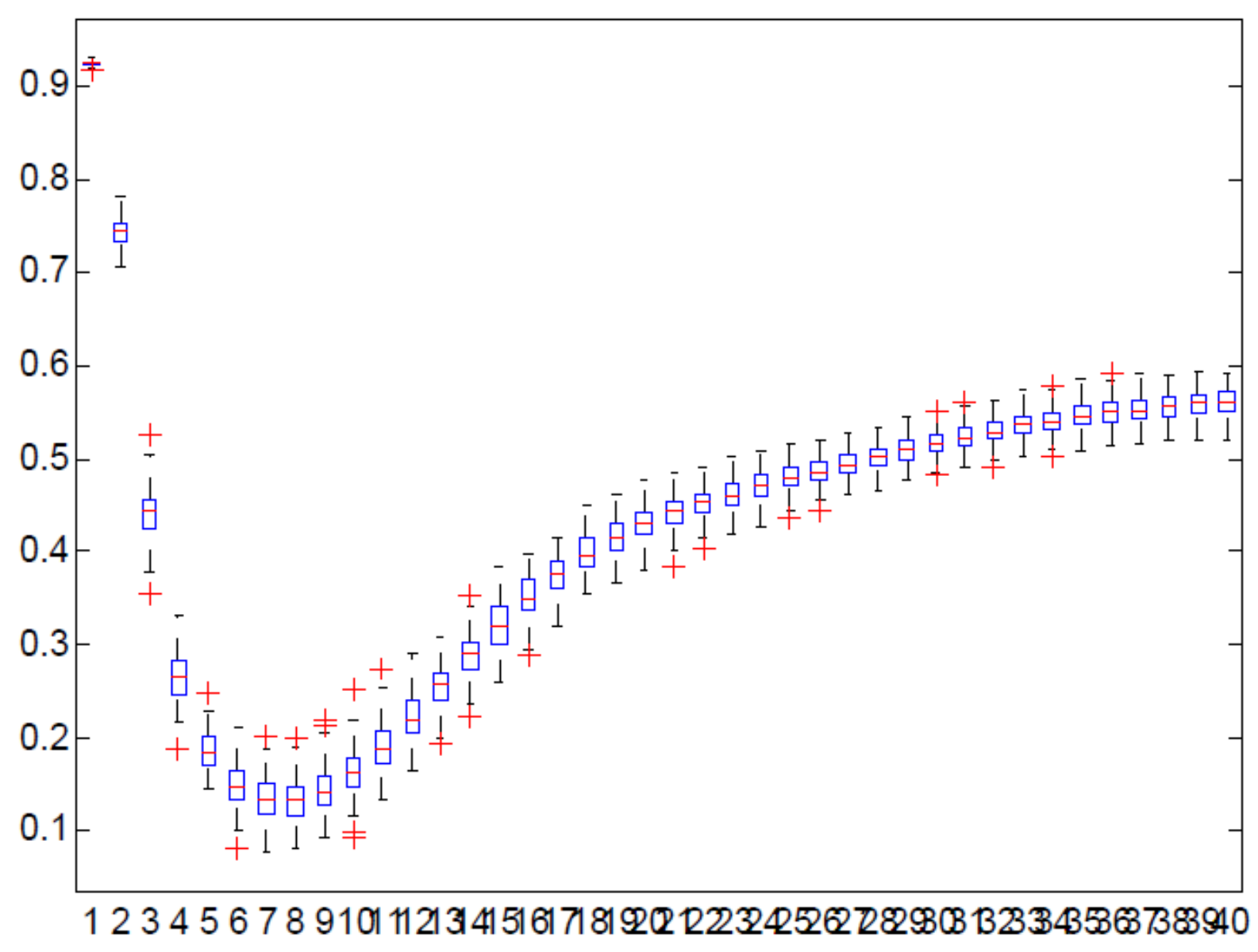} }
  \subfloat[][Hub graph]{\includegraphics[width=0.3\textwidth]{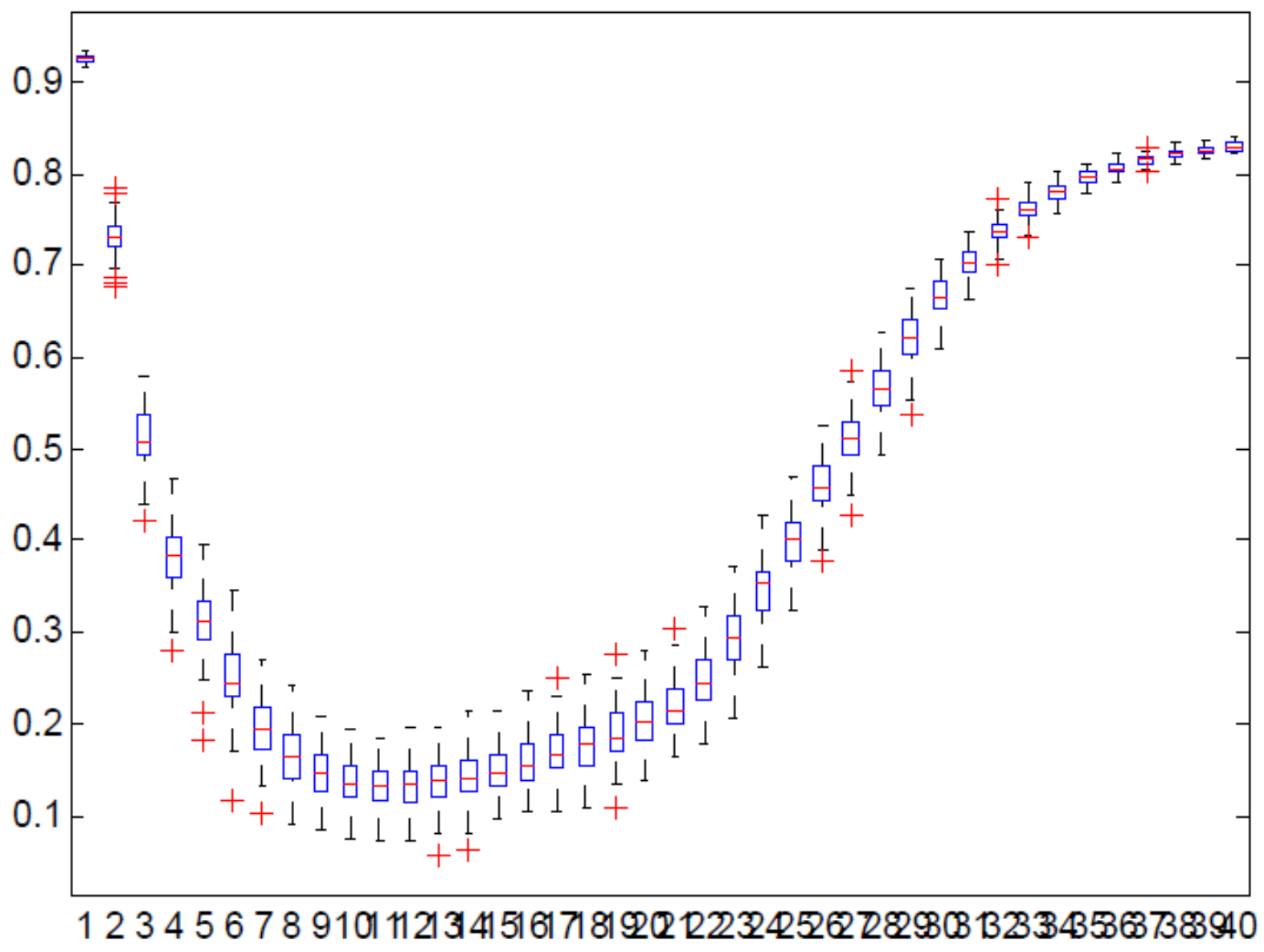} }
    \subfloat[][E-R graph]{\includegraphics[width=0.3\textwidth]{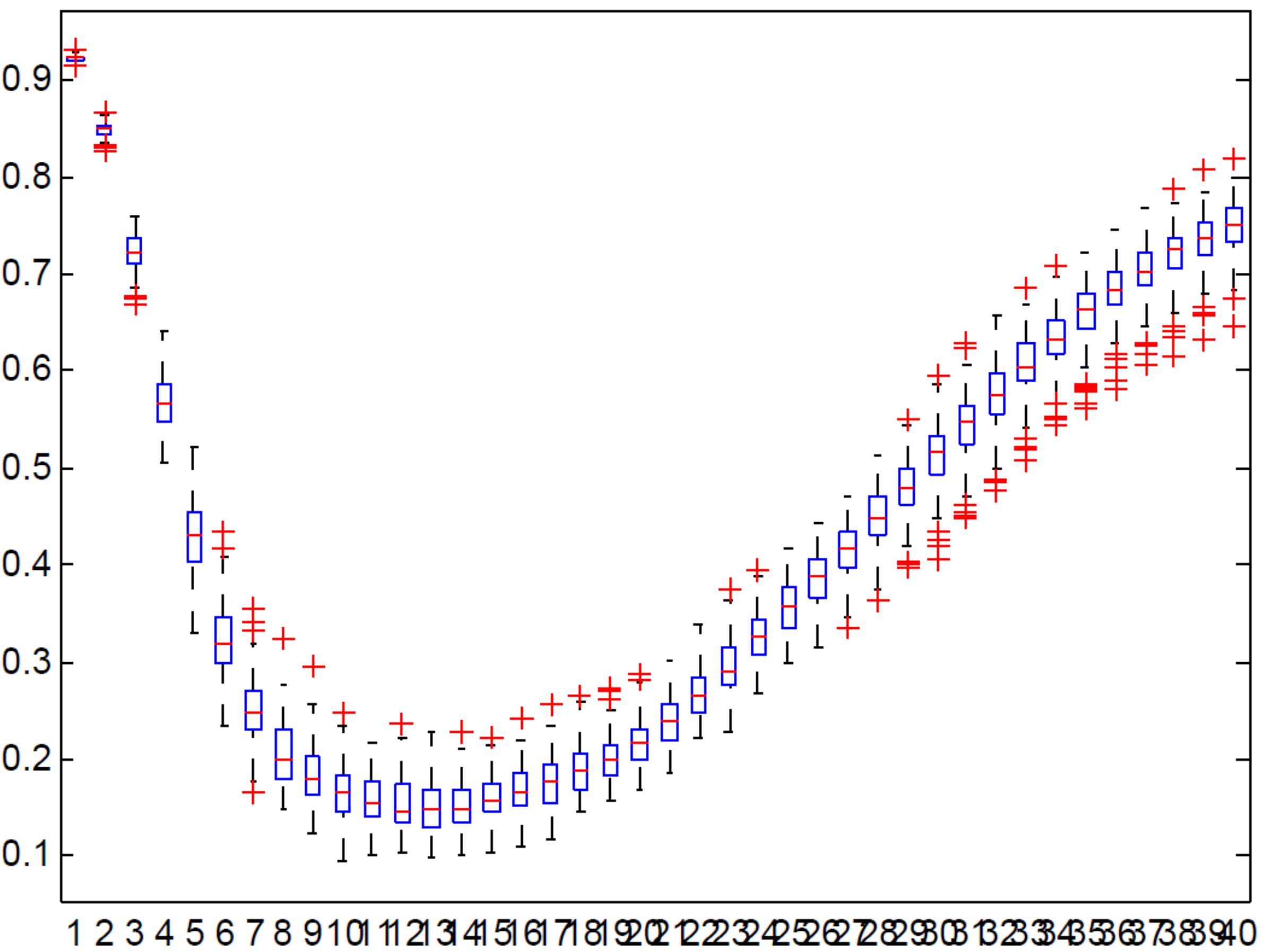} }
 \caption{FDP for $j=1,\ldots,40$ (GFC-Dantizg, $p=200$ and $\alpha=0.2$)}
 \label{fig:p50}
\end{figure}
\begin{figure}[htbp]
\centering
 \subfloat[][Band graph]{\includegraphics[width=0.3\textwidth]{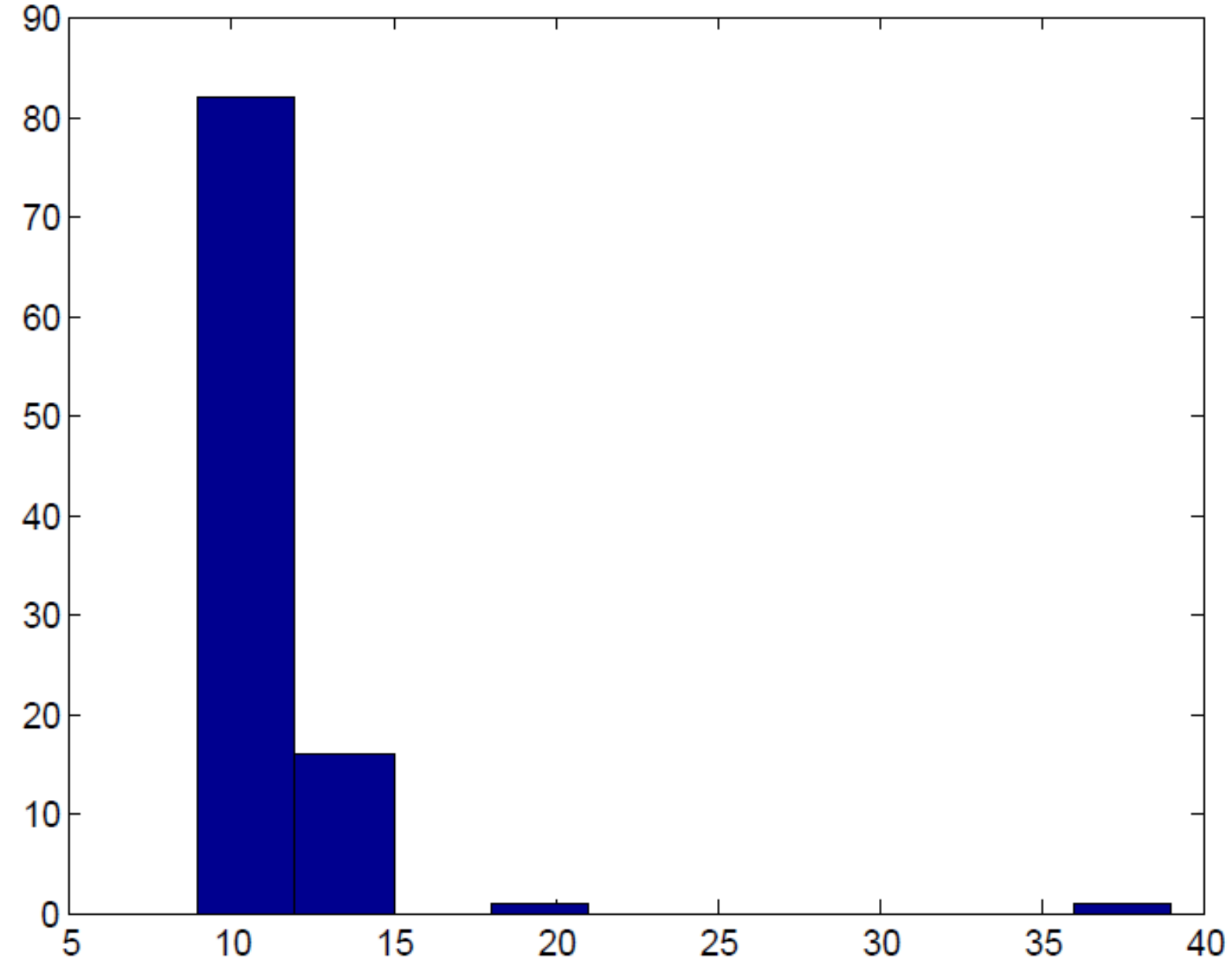} }
  \subfloat[][Hub graph]{\includegraphics[width=0.3\textwidth]{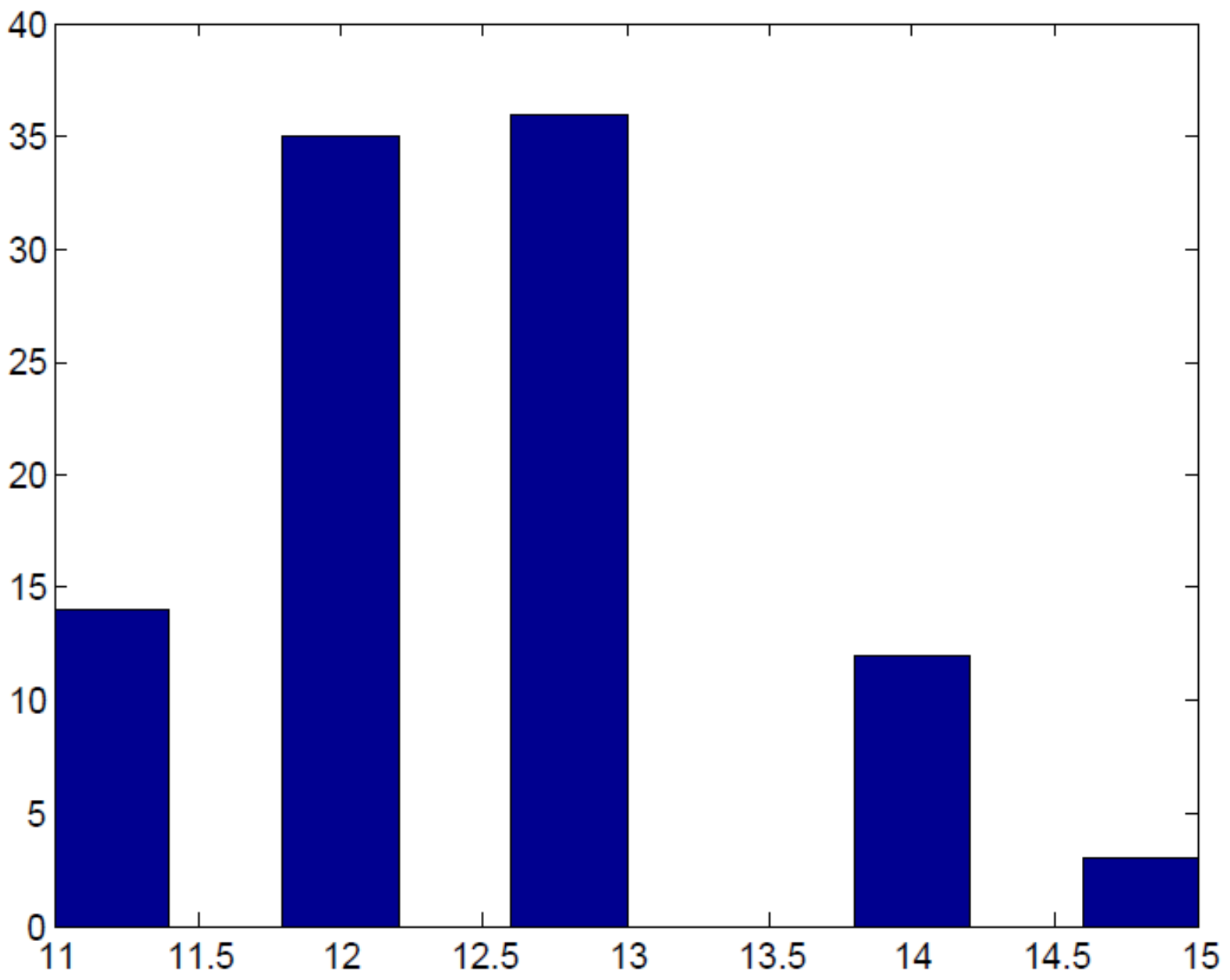} }
    \subfloat[][E-R graph]{\includegraphics[width=0.3\textwidth]{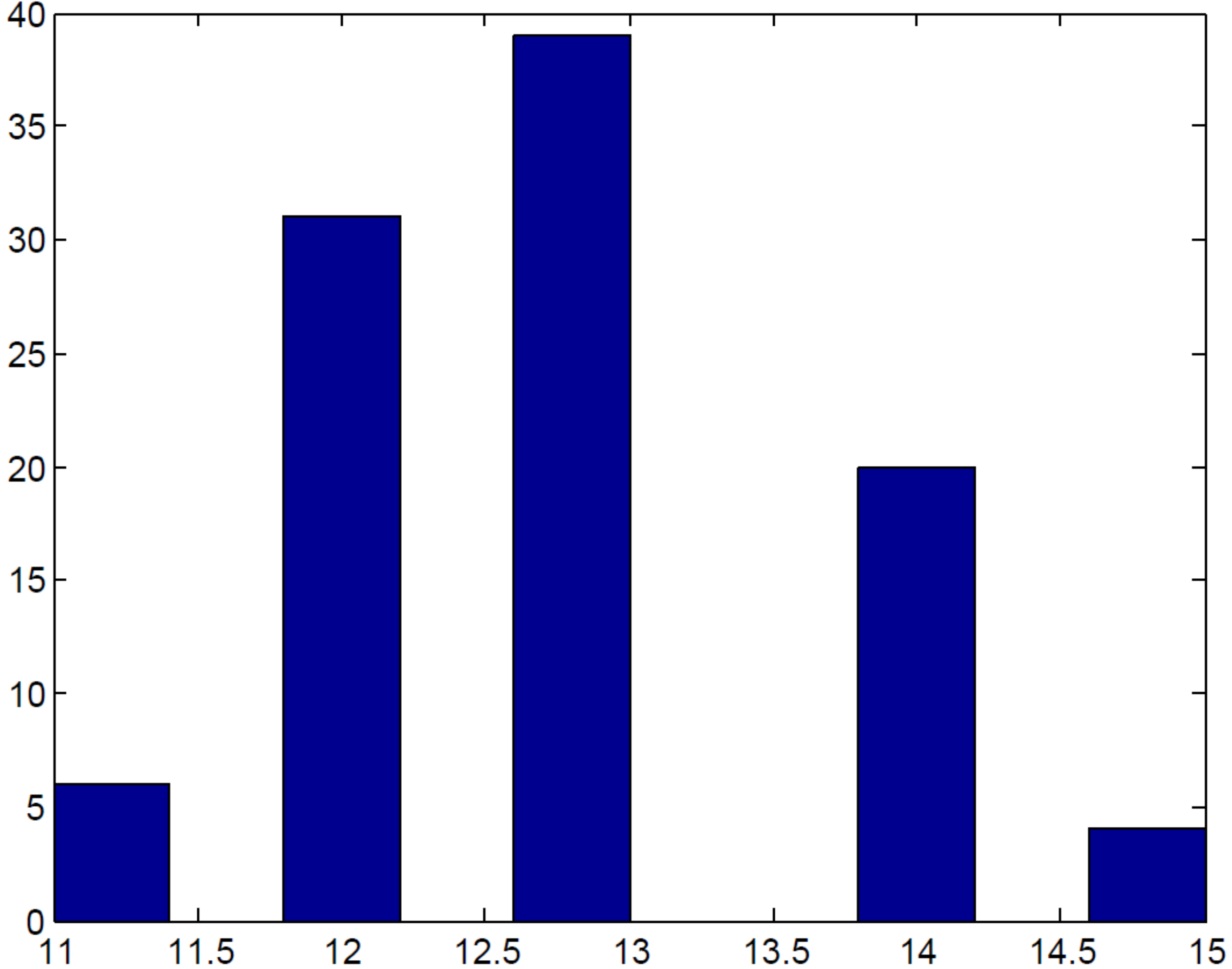} }
 \caption{Histogram for $\hat{j}$ (GFC-Dantizg, $p=200$ and $\alpha=0.2$)}
 \label{fig:p50}
\end{figure}

\begin{figure}[htbp]
\centering
 \subfloat[][Band graph]{\includegraphics[width=0.3\textwidth]{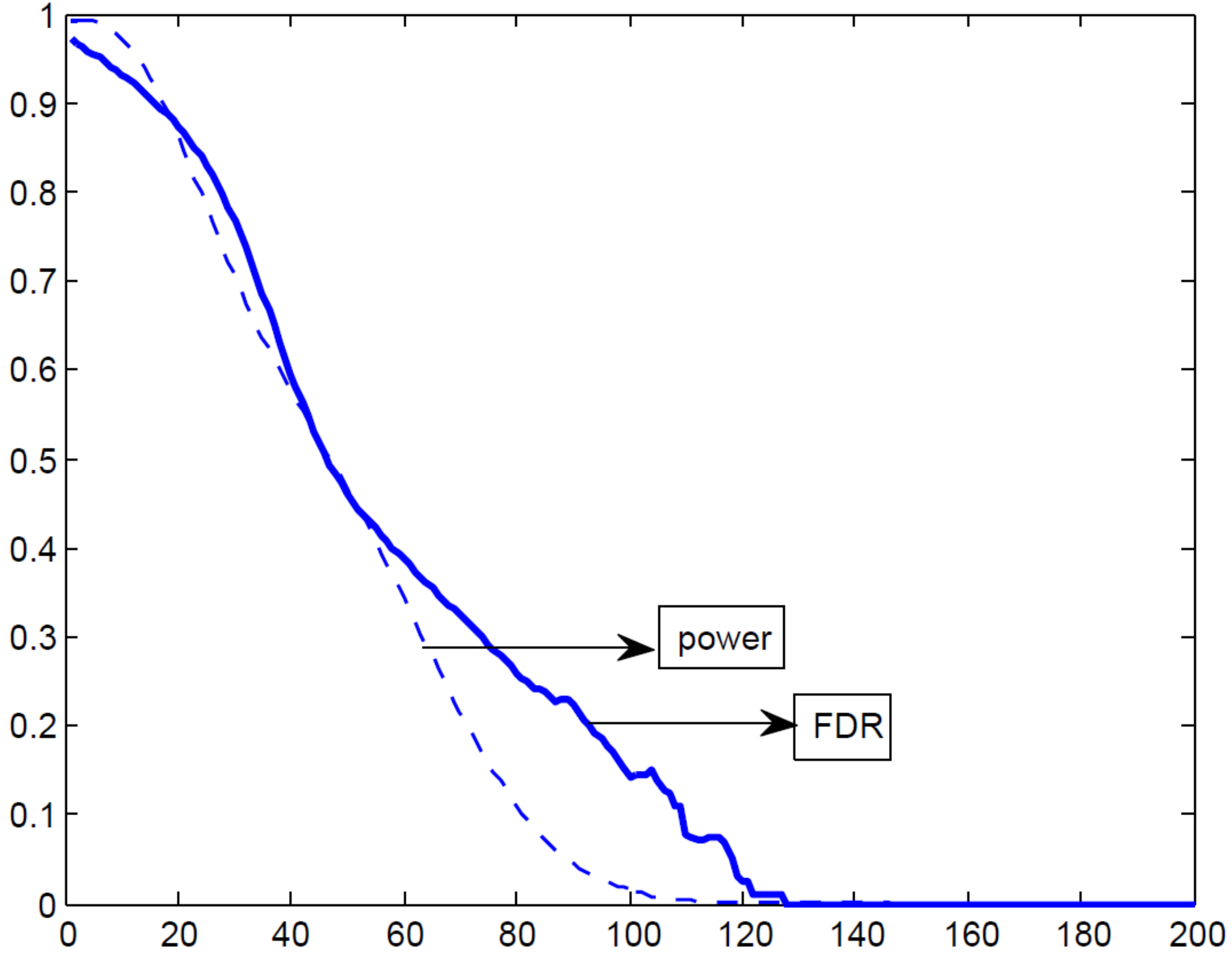} }
  \subfloat[][Hub graph]{\includegraphics[width=0.3\textwidth]{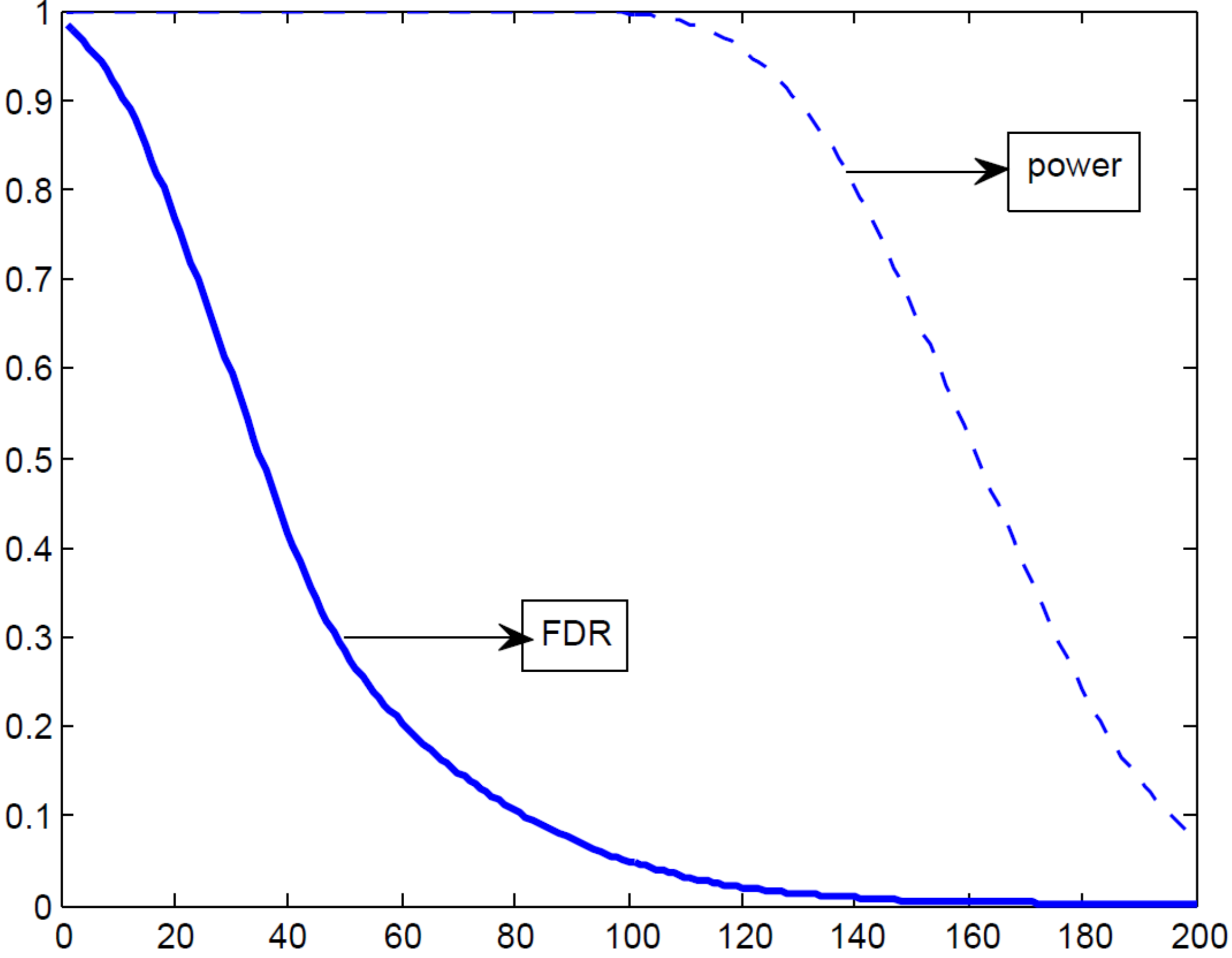} }
    \subfloat[][E-R graph]{\includegraphics[width=0.3\textwidth]{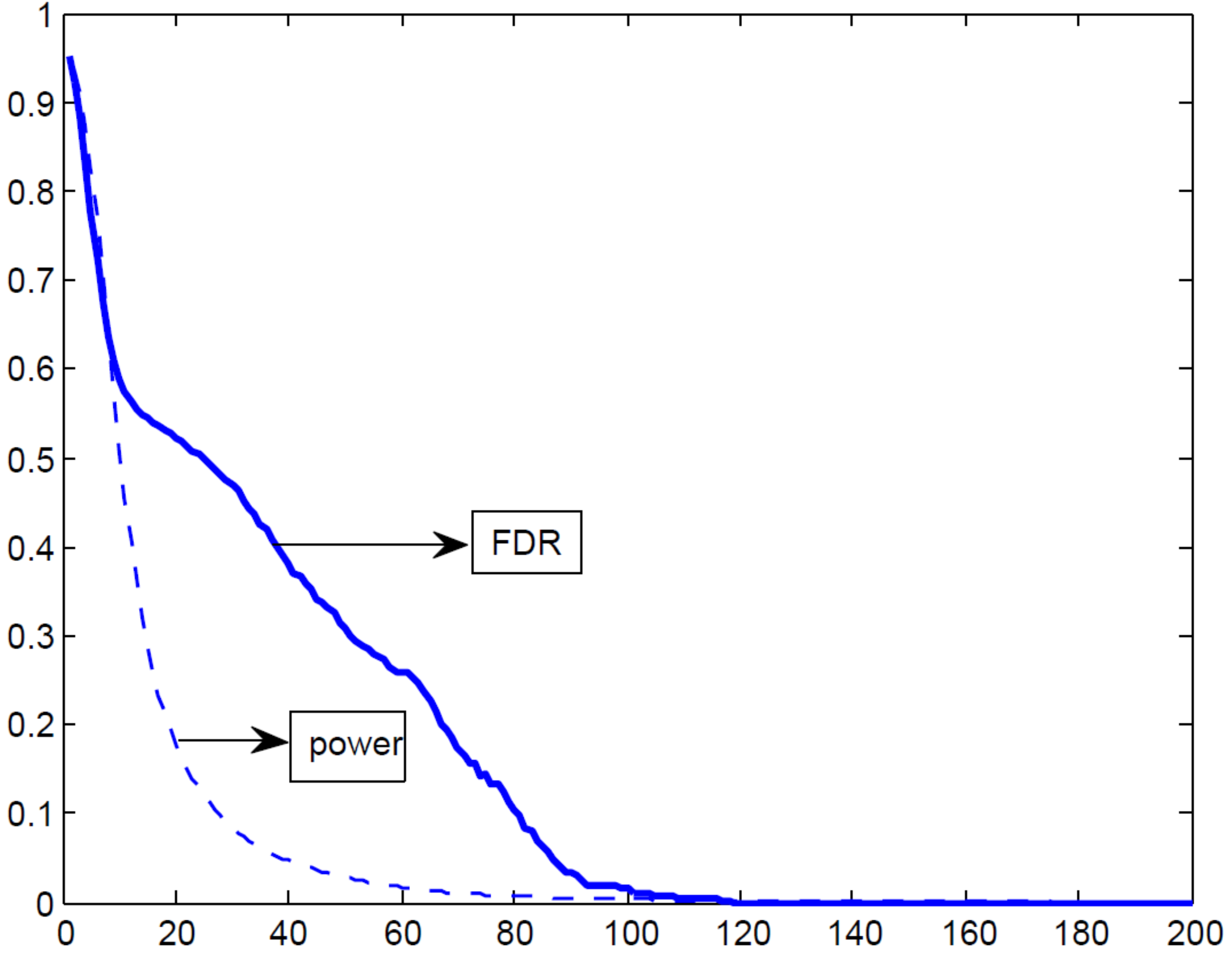} }
 \caption{FDR curve and power curve for graphical lasso ($p=200$)}
 \label{fig:p50}
\end{figure}

\section{Proof}

\subsection{Proof of Proposition 3.1}
Put $\tilde{\varepsilon}_{ki}=\varepsilon_{ki}-\bar{\varepsilon}_{i}$, where $(\bar{\varepsilon}_{1},\ldots,\bar{\varepsilon}_{p})^{'}=\bar{\va}$.
Recall the definitions of $\textbf{X}_{k,-j}$ and $\bar{\textbf{X}}_{-j}$ in Section 2.1. Note that
\begin{eqnarray}\label{p1}
\hat{\varepsilon}_{ki}\hat{\varepsilon}_{kj}&=&\tilde{\varepsilon}_{ki}\tilde{\varepsilon}_{kj}-
\tilde{\varepsilon}_{ki}(\textbf{X}_{k,-j}-\bar{\textbf{X}}_{-j})(\hat{\be}_{j}-\be_{j})\cr
& &-\tilde{\varepsilon}_{kj}(\textbf{X}_{k,-i}-\bar{\textbf{X}}_{-i})(\hat{\be}_{i}-\be_{i})\cr
& &+(\hat{\be}_{i}-\be_{i})^{'}(\textbf{X}_{k,-i}-\bar{\textbf{X}}_{-i})^{'}(\textbf{X}_{k,-j}-\bar{\textbf{X}}_{-j})(\hat{\be}_{j}-\be_{j}).
\end{eqnarray}
For the last term in (\ref{p1}), we have
\begin{eqnarray*}
|(\hat{\be}_{i}-\be_{i})^{'}\hat{\S}_{-i,-j}(\hat{\be}_{j}-\be_{j})|&\leq& |(\hat{\be}_{i}-\be_{i})^{'}(\hat{\S}_{-i,-j}-\S_{-i,-j})(\hat{\be}_{j}-\be_{j})|\cr
& &+|(\hat{\be}_{i}-\be_{i})^{'}\S_{-i,-j}(\hat{\be}_{j}-\be_{j})|.
\end{eqnarray*}
It is easy to show that, for any $M>0$, there exists $C>0$ such that
\begin{eqnarray}\label{p2}
\pr\Big{(}\max_{1\leq i<j\leq p}|\hat{\sigma}_{ij}-\sigma_{ij}|\geq C\sqrt{\log p/n}\Big{)}=O(p^{-M}).
\end{eqnarray}
Hence
\begin{eqnarray*}
\max_{i,j}|(\hat{\be}_{i}-\be_{i})^{'}(\hat{\S}_{-i,-j}-\S_{-i,-j})(\hat{\be}_{j}-\be_{j})|=O_{\pr}(a^{2}_{n1}(\log p/n)^{1/2}).
\end{eqnarray*}
Moreover,
\begin{eqnarray*}
|(\hat{\be}_{i}-\be_{i})^{'}\S_{-i,-j}(\hat{\be}_{j}-\be_{j})|= O_{\pr}(\lambda_{\max}(\S)|\hat{\be}_{i}-\be_{i}|_{2}^{2})
\end{eqnarray*}
uniformly in $1\leq i\leq j\leq p$. By the Cauchy-Schwarz inequality, we have
\begin{eqnarray*}
&&\Big{|}\frac{1}{n}\sum_{k=1}^{n}(\hat{\be}_{i}-\be_{i})^{'}(\textbf{X}_{k,-i}-\bar{\textbf{X}}_{-i})^{'}(\textbf{X}_{k,-j}-\bar{\textbf{X}}_{-j})(\hat{\be}_{j}-\be_{j})\Big{|}\cr
&&\quad \leq \max_{1\leq i\leq p}(\hat{\be}_{i}-\be_{i})^{'}\hat{\S}_{-i,-i}(\hat{\be}_{i}-\be_{i}).
\end{eqnarray*}
Combining the above arguments,
\begin{eqnarray*}
&&\Big{|}\frac{1}{n}\sum_{k=1}^{n}(\hat{\be}_{i}-\be_{i})^{'}(\textbf{X}_{k,-i}-\bar{\textbf{X}}_{-i})^{'}(\textbf{X}_{k,-j}-\bar{\textbf{X}}_{-j})(\hat{\be}_{j}-\be_{j})\Big{|}\cr
&&\quad=O_{\pr}(a_{n2}^{2}+a^{2}_{n1}(\log p/n)^{1/2}).
\end{eqnarray*}
We now estimate the second term on the right hand side of (\ref{p1}). For $1\leq i\leq j\leq p$, write
\begin{eqnarray*}
\tilde{\varepsilon}_{ki}(\textbf{X}_{k,-j}-\bar{\textbf{X}}_{-j})(\hat{\be}_{j}-\be_{j})&=&\tilde{\varepsilon}_{ki}(X_{ki}-\bar{X}_{i})
(\hat{\beta}_{i,j}-\beta_{i,j})I\{i\neq j\}\cr
& &+\sum_{l\neq i,j}\tilde{\varepsilon}_{ki}(X_{kl}-\bar{X}_{l})
(\hat{\beta}_{l,j}-\beta_{l,j}),
\end{eqnarray*}
where $\hat{\be}_{j}=(\hat{\beta}_{1,j},\ldots,\hat{\beta}_{p-1,j})^{'}$ and we set $\hat{\beta}_{p,j}=0$.
Recall that $\varepsilon_{ki}$ is independent with $\textbf{X}_{k,-j}$. Then it can be proved that, for any $M>0$, there exists $C>0$ such that
\begin{eqnarray*}
\pr\Big{(}\max_{1\leq i\leq p}\max_{1\leq l\leq p,l\neq i}\Big{|}\frac{1}{n}\sum_{k=1}^{n}\tilde{\varepsilon}_{ki}(X_{kl}-\bar{X}_{l})\Big{|}\geq C\sqrt{\frac{\log p}{n}}\Big{)}=O(p^{-M}).
\end{eqnarray*}
This implies that
\begin{eqnarray*}
\max_{1\leq i\leq j\leq p}\Big{|}\sum_{l\neq i,j}\frac{1}{n}\sum_{k=1}^{n}\tilde{\varepsilon}_{ki}(X_{kl}-\bar{X}_{l})
(\hat{\beta}_{l,j}-\beta_{l,j})\Big{|}=O_{\pr}(a_{n1}\sqrt{\log p/n}).
\end{eqnarray*}
A similar inequality holds for the third term on the right hand side of (\ref{p1}). Therefore,
\begin{eqnarray}\label{p4}
\frac{1}{n}\sum_{k=1}^{n}\hat{\varepsilon}_{ki}\hat{\varepsilon}_{kj}&=&\frac{1}{n}\sum_{k=1}^{n}\tilde{\varepsilon}_{ki}\tilde{\varepsilon}_{kj}
-\frac{1}{n}\sum_{k=1}^{n}\tilde{\varepsilon}_{ki}(X_{ki}-\bar{X}_{i})
(\hat{\beta}_{i,j}-\beta_{i,j})I\{i\neq j\}\cr
& &-\frac{1}{n}\sum_{k=1}^{n}\tilde{\varepsilon}_{kj}(X_{kj}-\bar{X}_{j})
(\hat{\beta}_{j-1,i}-\beta_{j-1,i})I\{i\neq j\}\cr
& &+O_{\pr}((a^{2}_{n1}+a_{n1})\sqrt{\log p/n}+a_{n2}^{2})
\end{eqnarray}
uniformly in $1\leq i\leq j\leq p$. By (\ref{reg}),
we have
\begin{eqnarray}\label{p3}
\frac{1}{n}\sum_{k=1}^{n}\tilde{\varepsilon}_{ki}(X_{ki}-\bar{X}_{i})=\frac{1}{n}\sum_{k=1}^{n}\tilde{\varepsilon}_{ki}^{2}
+\frac{1}{n}\sum_{k=1}^{n}\tilde{\varepsilon}_{ki}(\textbf{X}_{k,-i}-\bar{\textbf{X}}_{-i})\be_{i}.
\end{eqnarray}
By (C1), we have  $\Var(\textbf{X}_{k,-i}\be_{i})=(\sigma_{ii}\omega_{ii}-1)/\omega_{ii}\leq C$. It follows that
\begin{eqnarray*}
\pr\Big{(}\max_{1\leq i\leq p}\Big{|}\frac{1}{n}\sum_{k=1}^{n}\tilde{\varepsilon}_{ki}(\textbf{X}_{k,-i}-\bar{\textbf{X}}_{-i})\be_{i}\Big{|}\geq C\sqrt{\frac{\log p}{n}}\Big{)}=O(p^{-M}).
\end{eqnarray*}
By (\ref{p4}) and (\ref{p3}), we have, uniformly in $1\leq i\leq p$,
\begin{eqnarray}\label{p5}
\frac{1}{n}\sum_{k=1}^{n}\tilde{\varepsilon}_{ki}(X_{ki}-\bar{X}_{i})&=&\frac{1}{n}\sum_{k=1}^{n}\tilde{\varepsilon}_{ki}^{2}+O_{\pr}(\sqrt{\log p/n})\cr
&=&\frac{1}{n}\sum_{k=1}^{n}\hat{\varepsilon}_{ki}^{2}+O_{\pr}(\sqrt{\log p/n})\cr
& &+O_{\pr}((a^{2}_{n1}+a_{n1})\sqrt{\log p/n}+a_{n2}^{2}),
\end{eqnarray}
where the last equation follows from (\ref{p4}) with $i=j$. So, by (\ref{p4}), (\ref{p5}) and $\max_{i,j}|\hat{\beta}_{i,j}-\beta_{i,j}|= O_{\pr}(a_{n1})=o_{\pr}(1)$, for $1<i<j\leq p$,
\begin{eqnarray*}
\frac{1}{n}\sum_{k=1}^{n}\hat{\varepsilon}_{ki}\hat{\varepsilon}_{kj}&=&\frac{1}{n}\sum_{k=1}^{n}\tilde{\varepsilon}_{ki}\tilde{\varepsilon}_{kj}
-\frac{1}{n}\sum_{k=1}^{n}\hat{\varepsilon}_{ki}^{2}
(\hat{\beta}_{i,j}-\beta_{i,j})\cr
& &-\frac{1}{n}\sum_{k=1}^{n}\hat{\varepsilon}_{kj}^{2}
(\hat{\beta}_{j-1,i}-\beta_{j-1,i})\cr
& &+O_{\pr}((a^{2}_{n1}+a_{n1})\sqrt{\log p/n}+a_{n2}^{2}).
\end{eqnarray*}
By (\ref{p4}),  we have uniformly in $1\leq i\leq p$,
\begin{eqnarray}\label{a88}
\frac{1}{n}\sum_{k=1}^{n}\hat{\varepsilon}_{ki}^{2}&=&\frac{1}{n}\sum_{k=1}^{n}\tilde{\varepsilon}_{ki}^{2}
+O_{\pr}((a^{2}_{n1}+a_{n1})\sqrt{\log p/n}+a_{n2}^{2})
\end{eqnarray}
So, by (\ref{a88}) and $\max_{i,j}|\beta_{i,j}|\leq C$ for some constant $C>0$,
\begin{eqnarray}\label{a8}
&&\frac{1}{n}\sum_{k=1}^{n}\hat{\varepsilon}_{ki}\hat{\varepsilon}_{kj}+\frac{1}{n}\sum_{k=1}^{n}\hat{\varepsilon}_{ki}^{2}\hat{\beta}_{i,j}
+\frac{1}{n}\sum_{k=1}^{n}\hat{\varepsilon}_{kj}^{2}\hat{\beta}_{j-1,i}\cr
&&\quad=\frac{1}{n}\sum_{k=1}^{n}\tilde{\varepsilon}_{ki}\tilde{\varepsilon}_{kj}
+\frac{1}{n}\sum_{k=1}^{n}\hat{\varepsilon}_{ki}^{2}\beta_{i,j}+\frac{1}{n}\sum_{k=1}^{n}\hat{\varepsilon}_{kj}^{2}
\beta_{j-1,i}\cr
& &\quad\quad+O_{\pr}((a^{2}_{n1}+a_{n1})\sqrt{\log p/n}+a_{n2}^{2})\cr
&&\quad=-b_{nij}\frac{\omega_{ij}}{\omega_{ii}\omega_{jj}}+\frac{\sum_{k=1}^{n}(\varepsilon_{ki}\varepsilon_{kj}-\ep \varepsilon_{ki}\varepsilon_{kj}) }{n}\cr
&&\quad\quad+O_{\pr}\Big{(}a_{n1}\sqrt{\log p/n}+a_{n2}^{2}+\frac{\log p}{n}\Big{)}
\end{eqnarray}
uniformly in $1\leq i<j\leq p$. The proposition is proved by (C1) and the central limit theorem.\qed

\subsection{Proof of Theorem \ref{th1}}

To prove Theorem \ref{th1}, we need some lemmas.
 Let $\xi_{1},\ldots,\xi_{n}$ be independent and identically distributed $d$-dimensional random vectors with
mean zero. Let $G(t)=2-2\Phi(t)$ and define $|\cdot|_{(d)}$ by $|\z|_{(d)}=\min\{|z_{i}|; 1\leq i\leq d\}$ for $\z=(z_{1},\ldots,z_{d})^{'}$.

\begin{lemma}\label{le0} Suppose that $p\leq cn^{r}$ and $\ep|\xi_{1}|_{2}^{bdr+2+\epsilon}<\infty$ for some $c>0$, $r>0$, $b>0$ and $\epsilon>0$. Assume that
$\|\Cov(\xi_{1})-\I_{d}\|_{2}\leq C(\log p)^{-2-\gamma}$ for some $\gamma>0$. Then we have
\begin{eqnarray*}
\sup_{0\leq t\leq b\sqrt{\log p}}\Big{|}\frac{\pr(|\sum_{k=1}^{n}\xi_{k}|_{(d)}\geq t\sqrt{n})}{(G(t))^{d}}-1\Big{|}\leq C(\log p)^{-1-\gamma_{1}}.
\end{eqnarray*}
for  $\gamma_{1}=\min\{\gamma,1/2\}$.
\end{lemma}

\noindent{\bf Proof.} For $1\leq i\leq p$, put
\begin{eqnarray*}
&&\hat{\xi}_{i}=\xi_{i}I\{|\xi_{i}|_{2}\leq \sqrt{n}/(\log p)^{4}\}-\ep \xi_{i}I\{|\xi_{i}|_{2}\leq \sqrt{n}/(\log p)^{4}\},\cr
&&\tilde{\xi}_{i}=\xi_{i}-\hat{\xi}_{i}.
\end{eqnarray*}
We have
\begin{eqnarray*}
\pr(|\sum_{k=1}^{n}\xi_{k}|_{(d)}\geq t\sqrt{n})&\leq& \pr(|\sum_{k=1}^{n}\hat{\xi}_{k}|_{(d)}\geq t\sqrt{n}-\sqrt{n}/(\log p)^{2})\cr
& &+ \pr(|\sum_{k=1}^{n}\tilde{\xi}_{k}|_{2}\geq \sqrt{n}/(\log p)^{2}).
\end{eqnarray*}
Note that
\begin{eqnarray*}
\sum_{i=1}^{n}\ep |\xi_{i}|_{2}I\{|\xi_{i}|_{2}> \sqrt{n}/(\log p)^{4}\}=o(\sqrt{n}/(\log p)^{2}).
\end{eqnarray*}
We have by condition $\ep|\xi_{1}|_{2}^{bdr+2+\epsilon}<\infty$,
\begin{eqnarray*}
\pr(|\sum_{k=1}^{n}\tilde{\xi}_{k}|_{2}\geq \sqrt{n}/(\log p)^{2})\leq n\pr(|\xi_{1}|_{2}\geq \sqrt{n}/(\log p)^{4})\leq C(\log p)^{-3/2}(G(t))^{d}
\end{eqnarray*}
uniformly in $0\leq t\leq b\sqrt{\log p}$. Similarly, we have
\begin{eqnarray*}
\pr(|\sum_{k=1}^{n}\xi_{k}|_{(d)}\geq t\sqrt{n})&\geq& \pr(|\sum_{k=1}^{n}\hat{\xi}_{k}|_{(d)}\geq t\sqrt{n}+\sqrt{n}/(\log p)^{2})-C(\log p)^{-3/2}(G(t))^{d}.
\end{eqnarray*}
So it suffices to prove
\begin{eqnarray*}
\sup_{0\leq t\leq b\sqrt{\log p}}\Big{|}\frac{\pr(|\sum_{k=1}^{n}\hat{\xi}_{k}|_{(d)}\geq (t\pm(\log p)^{-2})\sqrt{n})}{(G(t))^{d}}-1\Big{|}\leq C(\log p)^{-1-\gamma_{1}}.
\end{eqnarray*}
By Theorem 1 in Za\"{\i}tsev (1987), we have
\begin{eqnarray*}
&&\pr(|\sum_{k=1}^{n}\hat{\xi}_{k}|_{(d)}\geq (t-(\log p)^{-2})\sqrt{n})\leq \pr(|\W|_{(d)}\geq t-2(\log p)^{-2})+c_{1,d}\exp(-c_{2,d}(\log p)^{2}),\cr
&&\pr(|\sum_{k=1}^{n}\hat{\xi}_{k}|_{(d)}\geq (t+(\log p)^{-2})\sqrt{n})\geq \pr(|\W|_{(d)}\geq t+2(\log p)^{-2})-c_{1,d}\exp(-c_{2,d}(\log p)^{2}),
\end{eqnarray*}
where $c_{1,d}$ and $c_{2,d}$ are positive constants depending only on $d$, $\W$ is a multivariate normal vector with mean zero and
covariance matrix $\Cov(\sum_{i=1}^{n}\hat{\xi}_{i}/\sqrt{n})$. We have
\begin{eqnarray*}
\|\Cov(\sum_{i=1}^{n}\hat{\xi}_{i}/\sqrt{n})-\I_{d}\|_{2}\leq C(\log p)^{-2-\gamma}.
\end{eqnarray*}
So it is easy to show that
\begin{eqnarray*}
\pr(|\W|_{(d)}\geq t-2(\log p)^{-2})\leq (1+C(\log p)^{-1-\gamma})(G(t))^{d}
\end{eqnarray*}
uniformly in $0\leq t\leq b\sqrt{\log p}$. By noting that $c_{1,d}\exp(-c_{2,d}(\log p)^{2})\leq C(\log p)^{-1-\gamma_{1}}(G(t))^{d}$ for $0\leq t\leq b\sqrt{\log p}$,
we obtain that
\begin{eqnarray*}
\pr(|\sum_{k=1}^{n}\hat{\xi}_{k}|_{(d)}\geq (t-(\log p)^{-2})\sqrt{n})\leq (1+C(\log p)^{-1-\gamma_{1}})(G(t))^{d}
\end{eqnarray*}
uniformly in $0\leq t\leq b\sqrt{\log p}$. Similarly, we can prove that
\begin{eqnarray*}
\pr(|\sum_{k=1}^{n}\hat{\xi}_{k}|_{(d)}\geq (t-(\log p)^{-2})\sqrt{n})\geq (1-C(\log p)^{-1-\gamma_{1}})(G(t))^{d}.
\end{eqnarray*}
This finishes the proof.\qed\\

Let  $\boldsymbol{\eta}_{k}=(\eta_{k1},\eta_{k2})^{'}$ are independent and identically distributed $2$-dimensional random vectors with
mean zero.

\begin{lemma} Suppose that $p\leq cn^{r}$ and $\ep|\boldsymbol{\eta}_{1}|_{2}^{2br+2+\epsilon}<\infty$ for some $c>0$, $r>0$, $b>0$ and $\epsilon>0$. Assume that
$\Var(\eta_{11})=\Var(\eta_{12})=1$ and $|\Cov(\eta_{11},\eta_{12})|\leq \delta$ for some $0\leq \delta<1$. Then we have
\begin{eqnarray*}
\pr\Big{(}|\sum_{k=1}^{n}\eta_{k1}|\geq t\sqrt{n},|\sum_{k=1}^{n}\eta_{k2}|\geq t\sqrt{n}\Big{)}\leq C(t+1)^{-2}\exp(-t^{2}/(1+\delta))
\end{eqnarray*}
uniformly for $0\leq t\leq b\sqrt{\log p}$, where $C$ only depends on $c,b,r,\epsilon,\delta$.
\end{lemma}

\noindent{\bf Proof.} The proof is similar to that of Lemma \ref{le0}. Actually, following the proof of Lemma \ref{le0},
we only need to prove
\begin{eqnarray}\label{a24}
\pr(|\W|_{(2)}\geq t-2(\log p)^{-2})\leq C(t+1)^{-2}\exp(-t^{2}/(1+\delta)),
\end{eqnarray}
where $\W$ is a two dimensional normal vector with mean zero and
covariance matrix $\Cov(\sum_{i=1}^{n}\hat{\boldsymbol{\eta}}_{i}/\sqrt{n})$ and
\begin{eqnarray*}
\hat{\boldsymbol{\eta}}_{i}=\boldsymbol{\eta}_{i}I\{|\boldsymbol{\eta}_{i}|_{2}\leq \sqrt{n}/(\log p)^{4}\}-\ep \boldsymbol{\eta}_{i}I\{|\boldsymbol{\eta}_{i}|_{2}\leq \sqrt{n}/(\log p)^{4}\}.
\end{eqnarray*}
We have
\begin{eqnarray*}
\|\Cov(\sum_{i=1}^{n}\hat{\boldsymbol{\eta}}_{i}/\sqrt{n})-\Cov(\boldsymbol{\eta}_{1})\|_{2}\leq C(\log p)^{-2-\gamma}.
\end{eqnarray*}
This, together with  Lemma 2 in Berman (1962) and some tedious calculations, implies (\ref{a24}).\qed

We now start to prove Theorem \ref{th1}. Let $\rho_{ij,\omega}=\omega_{ij}/\sqrt{\omega_{ii}\omega_{jj}}$.  Put
\begin{eqnarray*}
\sigma_{ii,\varepsilon}=\Var(\varepsilon_{i})\mbox{\quad and\quad} U_{ij}=\frac{\sum_{k=1}^{n}(\varepsilon_{ki}\varepsilon_{kj}-\ep \varepsilon_{ki}\varepsilon_{kj})}{\sqrt{n}\sigma^{1/2}_{ii,\varepsilon}\sigma^{1/2}_{jj,\varepsilon}}.
\end{eqnarray*}
Note that $\Var(\varepsilon_{ki}\varepsilon_{kj})=\sigma_{ii,\varepsilon}\sigma_{jj,\varepsilon}(1+\rho_{ij,\omega}^{2})$. By letting $b=4$ in Lemma \ref{le0},
\begin{eqnarray}\label{a9}
\max_{i,j}\sup_{0\leq t\leq 4\sqrt{\log p}}\Big{|}\frac{\pr(|U_{ij}|\geq t\sqrt{1+\rho_{ij,\omega}^{2}})}{G(t)}-1\Big{|}\leq C(\log p)^{-1-\gamma_{1}}.
\end{eqnarray}
By (\ref{p4}), it is easy to see that
\begin{eqnarray*}
\max_{1\leq i\leq p}|\hat{r}_{ii}-\sigma_{ii,\varepsilon}|=O_{\pr}\Big{(}\sqrt{\frac{\log p}{n}}\Big{)}.
\end{eqnarray*}
By (\ref{cd2}) and (\ref{a8}), we have
\begin{eqnarray*}
\max_{1\leq i<j\leq p}\Big{|}\sqrt{\frac{n}{\hat{r}_{ii}\hat{r}_{jj}}}(T_{ij}+\b_{nij}\frac{\omega_{ij}}{\omega_{ii}\omega_{jj}})-U_{ij}\Big{|}=o_{\pr}((\log p)^{-1/2}).
\end{eqnarray*}
This implies that
\begin{eqnarray*}
\pr\Big{(}\max_{1\leq i<j\leq p}\sqrt{\frac{n}{\hat{r}_{ii}\hat{r}_{jj}(1+\rho_{ij,\omega}^{2})}}|T_{ij}+\b_{nij}\frac{\omega_{ij}}{\omega_{ii}\omega_{jj}}|\geq
\Big{(}2-O\Big{(}\frac{1}{\log p}\Big{)}\Big{)}\sqrt{\log p}\Big{)}\rightarrow 0.
\end{eqnarray*}
Under the conditions of Theorem \ref{th1} and noting that $\max_{1\leq i\leq j\leq p}|\b_{nij}-1|=O_{\pr}(\sqrt{\log p/n})$, we have
\begin{eqnarray*}
\sum_{1\leq i<j\leq p}I\{|\hat{T}_{ij}|\geq 2\sqrt{\log p}\}\geq \max(c_{p},d_{p})
\end{eqnarray*}
with probability tending to one, where
\begin{eqnarray*}
c_{p}=\Big{(}\frac{1}{\sqrt{8\pi}\alpha}+\delta\Big{)}\sqrt{\log_{2}p}\mbox{\quad and\quad} d_{p}=\frac{1}{2}\max_{1\leq i\leq p}Card(\mathcal{A}_{i}(\gamma)).
\end{eqnarray*}
  Hence
\begin{eqnarray}\label{a26}
 \frac{(p^{2}-p)/2}{\max\{\sum_{1\leq i<j\leq p}I\{|\hat{T}_{ij}|\geq 2\sqrt{\log p}\},1\}}\leq \frac{p^{2}-p}{2}\frac{1}{\max(c_{p},d_{p})}
\end{eqnarray}
with probability tending to one. For $0<\theta<(1-\rho)/(1+\rho)$, let
\begin{eqnarray*}
\Lambda(\theta)=\{1\leq i\leq p: \exists j\neq i,~~s.t.~~\frac{|\omega_{ij}|}{\sqrt{\omega_{ii}\omega_{jj}}}\geq \theta\}.
\end{eqnarray*}
If Card$(\Lambda(\theta))\geq p/(\log p)^{6}$, then
\begin{eqnarray*}
\sum_{1\leq i<j\leq p}I\{|\hat{T}_{ij}|\geq 2\sqrt{\log p}\}\geq 2^{-1}p/(\log p)^{6}
\end{eqnarray*}
with probability tending to one and the upper bound in (\ref{a26}) can be replaced by $Cp(\log p)^{6}$. Set $d_{p}=\max_{1\leq i\leq p}Card(\mathcal{A}_{i}(\gamma))$. We let
\begin{eqnarray*}
b_{p}=G^{-1}\Big{(}p^{-2}\alpha\max\{c_{p},d_{p}\}\Big{)}\quad \mbox{and}\quad \theta_{1}=\theta
 \end{eqnarray*}
 if Card$(\Lambda(\theta))< p/(\log p)^{6}$;
 \begin{eqnarray*}
 b_{p}=\sqrt{2\log p+14\log_{2}p}\mbox{\quad and\quad} \theta_{1}=1
 \end{eqnarray*}
  if
Card$(\Lambda(\theta))\geq p/(\log p)^{6}$.
 Note that
\begin{eqnarray*}
1-\Phi(b_{p})\sim \frac{1}{\sqrt{2\pi}b_{p}}\exp(-b_{p}^{2}/2).
\end{eqnarray*}
Hence, by the definition of $\hat{t}$, we have $\pr(0\leq\hat{t}\leq b_{p})\rightarrow 1$. For $0\leq\hat{t}< 2\sqrt{\log p}$ and any $t<\hat{t}$, we have
\begin{eqnarray*}
\frac{G(t)(p^{2}-p)/2}{\max\{\sum_{1\leq i<j\leq p}I\{|\hat{T}_{ij}|\geq t\},1\}}>\alpha.
\end{eqnarray*}
This yields that for any $t<\hat{t}$
\begin{eqnarray*}
\frac{G(t)(p^{2}-p)/2}{\max\{\sum_{1\leq i<j\leq p}I\{|\hat{T}_{ij}|\geq \hat{t}\},1\}}>\alpha.
\end{eqnarray*}
By letting $t\rightarrow\hat{t}$, we obtain
\begin{eqnarray*}
\frac{G(\hat{t})(p^{2}-p)/2}{\max\{\sum_{1\leq i<j\leq p}I\{|\hat{T}_{ij}|\geq \hat{t}\},1\}}\geq\alpha.
\end{eqnarray*}
By the definition of infimum, there exists a sequence $t_{k}$ with $t_{k}\geq \hat{t}$, $t_{k}\rightarrow \hat{t}$ and
\begin{eqnarray*}
\frac{G(t_{k})(p^{2}-p)/2}{\max\{\sum_{1\leq i<j\leq p}I\{|\hat{T}_{ij}|\geq t_{k}\},1\}}\leq\alpha.
\end{eqnarray*}
It follows that
\begin{eqnarray*}
\frac{G(t_{k})(p^{2}-p)/2}{\max\{\sum_{1\leq i<j\leq p}I\{|\hat{T}_{ij}|\geq \hat{t}\},1\}}\leq\alpha.
\end{eqnarray*}
By letting $t_{k}\rightarrow \hat{t}$, we get
\begin{eqnarray*}
\frac{G(\hat{t})(p^{2}-p)/2}{\max\{\sum_{1\leq i<j\leq p}I\{|\hat{T}_{ij}|\geq \hat{t}\},1\}}\leq\alpha.
\end{eqnarray*}
Hence, when $0\leq\hat{t}<2\sqrt{\log p}$,
\begin{eqnarray*}
\frac{G(\hat{t})(p^{2}-p)/2}{\max\{\sum_{1\leq i<j\leq p}I\{|\hat{T}_{ij}|\geq \hat{t}\},1\}}=\alpha.
\end{eqnarray*}
To prove Theorem \ref{th1}, by  $\pr(0\leq\hat{t}\leq b_{p})\rightarrow 1$, it is enough to show that
\begin{eqnarray}\label{p6}
\sup_{0\leq t\leq b_{p}}\Big{|}\frac{\sum_{(i,j)\in\mathcal{H}_{0}}I\{|\hat{T}_{ij}|\geq t\}}{q_{0}G(t)}-1\Big{|}\rightarrow 0
\end{eqnarray}
in probability, where $q_{0}=Card(\mathcal{H}_{0})$.
To prove (\ref{p6}), we need the following lemma.

\begin{lemma}\label{le3} Suppose that for any $\varepsilon>0$,
\begin{eqnarray}\label{v1-1}
\sup_{0\leq t\leq b_{p}}\pr\Big{(}\Big{|}\frac{\sum_{(i,j)\in\mathcal{H}_{0}}[I\{|U_{ij}|\geq t\}-\pr(|U_{ij}|\geq t)]}{2q_{0}(1-\Phi(t))}\Big{|}\geq \varepsilon\Big{)}=o(1),
\end{eqnarray}
and
\begin{eqnarray}\label{v1}
&&\int_{0}^{b_{p}}\pr\Big{(}\Big{|}\frac{\sum_{(i,j)\in\mathcal{H}_{0}}[I\{|U_{ij}|\geq t\}-\pr(|U_{ij}|\geq t)]}{2q_{0}(1-\Phi(t))}\Big{|}\geq \varepsilon\Big{)} dt=o(v_{p}),
\end{eqnarray}
where $v_{p}=1/\sqrt{(\log p)(\log_{4}p)^{2}}$. Then (\ref{p6}) holds.
\end{lemma}

Let's first finish the proof of Theorem \ref{th1}. By Lemma \ref{le3}, it suffices to prove (\ref{v1-1}) and (\ref{v1}). Define
\begin{eqnarray*}
&&\mathcal{S}_{1}=\left\{
\begin{array}{c c}
\{(i,j): i\in\Lambda(\theta), j\geq i\}& \mbox{if Card} (\Lambda(\theta))< p/(\log p)^{6}\\
\emptyset&\mbox{if Card} (\Lambda(\theta))\geq p/(\log p)^{6}
\end{array}\right.,\cr
&&\mathcal{S}_{2}=\{(i,j): 1\leq i\leq p, j\in \mathcal{A}_{i}(\gamma)\},\cr
&&\mathcal{H}_{01}=\mathcal{H}_{0}\cap \{\mathcal{S}_{1}\cup \mathcal{S}_{2}\} ,\quad \mathcal{H}_{02}=\mathcal{H}_{0}\cap  \{\mathcal{S}_{1}\cup \mathcal{S}_{2}\}^{c}.
\end{eqnarray*}
Recall that $q_{0}\geq cp^{2}$.
Thus, by (\ref{a9}),
\begin{eqnarray}\label{t0}
\ep\Big{|}\frac{\sum_{(i,j)\in\mathcal{H}_{01}}[I\{|U_{ij}|\geq t\}-\pr(|U_{ij}|\geq t)]}{q_{0}G(t)}\Big{|}&\leq& C\frac{(p^{1+\rho}+p^{2}/(\log p)^{6})G(t)}{p^{2}G(t)}\cr
&=&O((\log p)^{-6})
\end{eqnarray}
uniformly for $0\leq t\leq 2\sqrt{\log p}$.
Note that
\begin{eqnarray}\label{t1}
&&\ep\Big{[}\frac{\sum_{(i,j)\in\mathcal{H}_{02}}\Big{\{}I\{|U_{ij}|\geq t\}-\pr(|U_{ij}|\geq t)\Big{\}}}{q_{0}G(t)}\Big{]}^{2}\cr
&&=\frac{\sum_{(i,j)\in\mathcal{H}_{02}}\sum_{(k,l)\in\mathcal{H}_{02}}
\{\pr(|U_{ij}|\geq t,|U_{kl}|\geq t)-\pr(|U_{ij}|\geq t)\pr(|U_{kl}|\geq t)\}
}{q^{2}_{0}G^{2}(t)}.
\end{eqnarray}
We next split the set $\mathcal{H}_{02}$ into two subsets as in Cai, Liu and Xia (2013). Let $G_{abcd}=(V_{abcd},E_{abcd})$ be a graph, where $V_{abcd}=\{a,b,c,d\}$ is the set of vertices and $E_{abcd}$ is the set of edges.
There is an edge between $i\neq j\in \{a,b,c,d\}$ if and only if  $|\omega_{ij}|\geq (\log p)^{-2-\gamma}$.
 If the number of different vertices in $V_{abcd}$ is 3, then we call $G_{abcd}$ as a three vertices graph (3-G).
 Similarly, $G_{abcd}$ is a four vertices graph (4-G)
if the number of different vertices in $V_{abcd}$ is 4. A vertex in $G_{abcd}$ is said to be {\em isolated} if  there is no edge connected to it.
Note that for any $(i,j)\in\mathcal{H}_{02}$, $(k,l)\in\mathcal{H}_{02}$ and $(i,j)\neq (k,l)$, $G_{ijkl}$ is 3-G or 4-G. We say a graph $\mathcal{G}:=G_{ijkl}$ satisfy
$(\star)$ if
\begin{eqnarray*}
(\star):&&\mbox{If $\mathcal{G}$ is 4-G, then there is at least one isolated vertex in $\mathcal{G}$;}\cr
&&\quad\mbox{otherwise  $\mathcal{G}$ is 3-G and $E_{ijkl}=\emptyset$.}
\end{eqnarray*}
For any $G_{ijkl}$ satisfying $(\star)$,
\begin{eqnarray}\label{p20}
|\ep [\varepsilon_{i}\varepsilon_{j}\varepsilon_{k}\varepsilon_{l}]|=O((\log p)^{-2-\gamma}),
\end{eqnarray}
where $O(1)$ is uniformly for $i,j,k,l$. By the above definition, we  further divide the indices set in (\ref{t1}) into
\begin{eqnarray*}
\mathcal{H}_{020}&=&\{(i,j)\in\mathcal{H}_{02},(k,l)\in\mathcal{H}_{02}:(i,j)=(k.l)\};\cr
\mathcal{H}_{021}&=&\{(i,j)\in\mathcal{H}_{02},(k,l)\in\mathcal{H}_{02}:(i,j)\neq(k.l),\mbox{$\mathcal{G}_{ijkl}$ satisfies $(\star)$.}\};\cr
\mathcal{H}_{022}&=&\{(i,j)\in\mathcal{H}_{02},(k,l)\in\mathcal{H}_{02}: (i,j)\neq(k.l),\mbox{$\mathcal{G}_{ijkl}$ does not satisfy $(\star)$.}\}.
\end{eqnarray*}
For the indices in $\mathcal{H}_{020}$, we have by  (\ref{a9}),
\begin{eqnarray}\label{a23}
\Big{|}\frac{\sum_{\{(i,j),(k,l)\}\in\mathcal{H}_{020}}
\{\pr(|U_{ij}|\geq t,|U_{kl}|\geq t)-\pr(|U_{ij}|\geq t)\pr(|U_{kl}|\geq t)\}
}{q^{2}_{0}G^{2}(t)}\Big{|}\leq \frac{C}{p^{2}G(t)}.
\end{eqnarray}
It is easy to show that Card($\mathcal{H}_{022}$)$\leq Cp^{2}d_{p}^{2}$. We say the graph $\mathcal{G}_{ijkl}$ is $a$G-$b$E if $G_{ijkl}$ is $a$-G and there are
$b$ edges in $E_{ijkl}$ for $a=3,4$ and $b=0,1,2,3,4$. Note that for any $(i,j)\in\mathcal{H}_{0}$, the vertices $i$ and $j$ are not connected.  So we can divide $\mathcal{H}_{022}$ into two parts:
\begin{eqnarray*}
\mathcal{H}_{022,1}&=&\{\{(i,j),(k,l)\}\in \mathcal{H}_{022}: \mathcal{G}_{ijkl}\mbox{\quad is 3G-1E or 4G-2E}\},\cr
\mathcal{H}_{022,2}&=&\{\{(i,j),(k,l)\}\in \mathcal{H}_{022}: \mathcal{G}_{ijkl}\mbox{\quad is 4G-3E or 4G-4E}\}.
\end{eqnarray*}
It can be shown that Card$(\mathcal{H}_{022,2})=O(pd_{p}^{3})$ and  Card$(\mathcal{H}_{022,1})=O(p^{2}d_{p}^{2})$.
Then, by (\ref{a9}),
\begin{eqnarray}\label{t3-3}
\Big{|}\frac{\sum_{\{(i,j),(k,l)\}\in\mathcal{H}_{022,2}}
\{\pr(|U_{ij}|\geq t,|U_{kl}|\geq t)-\pr(|U_{ij}|\geq t)\pr(|U_{kl}|\geq t)\}
}{q^{2}_{0}G^{2}(t)}\Big{|}\leq \frac{Cd_{p}^{3}}{p^{3}G(t)}.
\end{eqnarray}
It remains for us to estimate the terms in $\mathcal{H}_{022,1}$ and $\mathcal{H}_{021}$. To this end, we need the following lemma.

\begin{lemma}\label{le1} We have
\begin{eqnarray}\label{a21}
\max_{\{(i,j),(k,l)\}\in\mathcal{H}_{021}}\pr\Big{(}|U_{ij}|\geq t,|U_{kl}|\geq t\Big{)}
=(1+A_{n})G^{2}(t)
\end{eqnarray}
and
\begin{eqnarray}\label{a22}
\max_{\{(i,j),(k,l)\}\in\mathcal{H}_{022,1}}\pr\Big{(}|U_{ij}|\geq t,|U_{kl}|\geq t\Big{)}\leq C(t+1)^{-1}\exp(-t^{2}/(1+\theta_{1}))
\end{eqnarray}
uniformly in $0\leq t\leq b_{p}$, where $A_{n}\leq C(\log p)^{-1-\gamma_{1}}$.

\end{lemma}

\noindent{\bf Proof.}
It can be proved that, uniformly for $\{(i,j),(k,l)\}\in\mathcal{H}_{021}$,
\begin{eqnarray*}
\Big{\|}\text{Corr}((\varepsilon_{ij},\varepsilon_{kl}))-\I_{2}\Big{\|}_{2}=O((\log p)^{-2-\gamma}),
\end{eqnarray*}
and uniformly for $\{(i,j),(k,l)\}\in\mathcal{H}_{022,1}$,
\begin{eqnarray*}
|\text{Corr}(\varepsilon_{ij},\varepsilon_{kl})|\leq \theta_{1}+O((\log p)^{-2-\gamma}).
\end{eqnarray*}
The proof is complete by Lemma \ref{le0}. \qed

By Lemma \ref{le1}, we have
\begin{eqnarray}\label{t6}
\Big{|}\frac{\sum_{\{(i,j),(k,l)\}\in\mathcal{H}_{021}}
\{\pr(U_{ij}\geq t,U_{kl}\geq t)-\pr(U_{ij}\geq t)\pr(U_{kl}\geq t)\}
}{q^{2}_{0}G^{2}(t)}\Big{|}\leq C(\log p)^{-1-\gamma_{1}}
\end{eqnarray}
and
\begin{eqnarray}\label{t7}
\Big{|}\frac{\sum_{\{(i,j),(k,l)\}\in\mathcal{H}_{022,1}}
\{\pr(U_{ij}\geq t,U_{kl}\geq t)-\pr(U_{ij}\geq t)\pr(U_{kl}\geq t)\}
}{q^{2}G^{2}(t)}\Big{|}\leq C{p^{-2}d_{p}^{2}[G(t)]^{-\frac{2\theta_{1}}{1+\theta_{1}}}}.
\end{eqnarray}
Combining (\ref{t0}), (\ref{t3-3}), (\ref{t6}), (\ref{t7}) and the fact $d_{p}=O(p^{\rho})$, we prove (\ref{v1}). The proof of (\ref{v1-1}) is exactly the same with that of (\ref{v1}) and hence is omitted.\qed\\

\noindent{\bf Proof of Lemma \ref{le3}.} Recall the definition of $b_{p}$ in the proof of Theorem \ref{th1}. Let $0=t_{0}<t_{1}<\cdots<t_{m}=b_{p}$ satisfy $t_{i}-t_{i-1}=v_{p}$ for $1\leq i\leq m-1$ and $t_{m}-t_{m-1}\leq v_{p}$. So $m\sim b_{p}/v_{p}$. For any $t_{j-1}\leq t\leq t_{j}$, we have
\begin{eqnarray}\label{a6}
\frac{\sum_{(i,j)\in\mathcal{H}_{0}}I\{|\hat{T}_{ij}|\geq t\}}{q_{0}G(t)}\leq\frac{\sum_{(i,j)\in\mathcal{H}_{0}}I\{|\hat{T}_{ij}|\geq t_{j-1}\}}{q_{0}G(t_{j-1})}\frac{G(t_{j-1})}{G(t_{j})}
\end{eqnarray}
and
\begin{eqnarray}\label{a7}
\frac{\sum_{(i,j)\in\mathcal{H}_{0}}I\{|\hat{T}_{ij}|\geq t\}}{q_{0}G(t)}\geq\frac{\sum_{(i,j)\in\mathcal{H}_{0}}I\{|\hat{T}_{ij}|\geq t_{j}\}}{q_{0}G(t_{j})}\frac{G(t_{j})}{G(t_{j-1})}.
\end{eqnarray}
In view of (\ref{a6}) and (\ref{a7}), we only need to prove
\begin{eqnarray*}
\max_{0\leq j\leq m}\Big{|}\frac{\sum_{(i,j)\in\mathcal{H}_{0}}[I\{|\hat{T}_{ij}|\geq t_{j}\}-G(t_{j})]}{q_{0}G(t_{j})}\Big{|}\rightarrow 0
\end{eqnarray*}
in probability.
We have
\begin{eqnarray*}
\max_{1\leq i<j\leq p}|\hat{T}_{ij}-U_{ij}|=O_{\pr}(a_{n1}\sqrt{\log p}+\sqrt{n}a_{n2}^{2}+(\log p)/\sqrt{n}).
\end{eqnarray*}
Since
\begin{eqnarray*}
\frac{G(t+o(\sqrt{1/\log p}))}{G(t)}=1+o(1)
\end{eqnarray*}
uniformly in $0\leq t\leq 2\sqrt{\log p}$, by (\ref{cd2}), it suffices to show that
\begin{eqnarray*}
\max_{0\leq j\leq m}\Big{|}\frac{\sum_{(i,j)\in\mathcal{H}_{0}}[I\{|U_{ij}|\geq t_{j}\}-G(t_{j})]}{q_{0}G(t_{j})}\Big{|}\rightarrow 0
\end{eqnarray*}
in probability.
We have
\begin{eqnarray*}
&&\pr\Big{(}\max_{1\leq j\leq m}\Big{|}\frac{\sum_{(i,j)\in\mathcal{H}_{0}}[I\{|U_{ij}|\geq t_{j}\}-G(t_{j})]}{q_{0}G(t_{j})}\Big{|}\geq \varepsilon\Big{)}\cr
&&\quad\leq\sum_{j=1}^{m}\pr\Big{(}\Big{|}\frac{\sum_{(i,j)\in\mathcal{H}_{0}}[I\{|U_{ij}|\geq t_{j}\}-G(t_{j})]}{q_{0}G(t_{j})}\Big{|}\geq \varepsilon\Big{)}\cr
&&\quad\leq \frac{1}{v_{p}}\int_{0}^{b_{p}}\pr\Big{(}\frac{\sum_{(i,j)\in\mathcal{H}_{0}}I\{|U_{ij}|\geq t\}}{q_{0}G(t)}\geq 1+\varepsilon/2\Big{)}dt\cr
&&\quad\quad+\frac{1}{v_{p}}\int_{0}^{b_{p}}\pr\Big{(}\frac{\sum_{(i,j)\in\mathcal{H}_{0}}I\{|U_{ij}|\geq t\}}{q_{0}G(t)}\leq 1-\varepsilon/2\Big{)}dt\cr
&&\quad\quad+\sum_{j=m-1}^{m}\pr\Big{(}\Big{|}\frac{\sum_{(i,j)\in\mathcal{H}_{0}}[I\{|U_{ij}|\geq t_{j}\}-G(t_{j})]}{q_{0}G(t_{j})}\Big{|}\geq \varepsilon\Big{)}.
\end{eqnarray*}
So it suffices to prove
\begin{eqnarray*}
\int_{0}^{b_{p}}\pr\Big{(}\Big{|}\frac{\sum_{(i,j)\in\mathcal{H}_{0}}I\{|U_{ij}|\geq t\}-G(t)}{q_{0}G(t)}\Big{|}\geq \varepsilon\Big{)}dt=o(v_{p})
\end{eqnarray*}
and
\begin{eqnarray*}
\sum_{k=m-1}^{m}\pr\Big{(}\Big{|}\frac{\sum_{(i,j)\in\mathcal{H}_{0}}[I\{|U_{ij}|\geq t_{k}\}-G(t_{k})]}{q_{0}G(t_{k})}\Big{|}\geq \varepsilon\Big{)}=o(1),
\end{eqnarray*}
which are the conditions of  Lemma \ref{le3}. \qed

\subsection{Proof of Propositions \ref{pro2} and \ref{pro3-3}}

\noindent{\bf Proof of Proposition \ref{pro2}.}
We first show that the true $\be_{i}$ belongs to the region
\begin{eqnarray}\label{a25}
|\D_{i}^{-1/2}\hat{\S}_{-i,-i}\be_{i}-\D_{i}^{-1/2}\hat{\a}|_{\infty}\leq \lambda_{ni}(2)
\end{eqnarray}
with probability tending to one. Without loss of generality, we assume $\ep \X_{k}=0$. It suffices to prove that
\begin{eqnarray*}
&&\Big{|}\frac{1}{n}\sum_{k=1}^{n}(X_{kj}-\bar{X}_{j})\Big{\{}\sum_{l\neq i}(X_{kl}-\bar{X}_{l})\beta_{l}-X_{ki}+\bar{X}_{i}\Big{\}}\Big{|}\cr
&&=
\Big{|}\frac{1}{n}\sum_{k=1}^{n}(X_{kj}-\bar{X}_{j})\varepsilon_{ki}\Big{|}\leq \sqrt{\hat{\sigma}_{jj}}\lambda_{ni}(2),
\end{eqnarray*}
uniformly in $1\leq i\neq j\leq p$, with probability tending to one. By the independence between $\{\varepsilon_{ki}\}$ and
 $\{X_{k,j},j\neq i\}$, we have
 \begin{eqnarray*}
 \pr\Big{(}\max_{i\neq j}\frac{1}{\sqrt{n\hat{\sigma}_{jj}\Var(\varepsilon_{i})}}\Big{|}\sum_{k=1}^{n}(X_{kj}-\bar{X}_{j})\varepsilon_{ki}\Big{|}\geq (2+O((\log p)^{-1/2})\sqrt{\log p}\Big{)}\leq C(\log p)^{-1/2}.
 \end{eqnarray*}
 Since $\Var(\varepsilon_{i})=1/\omega_{ii}\leq \sigma_{ii}$, we prove (\ref{a25}). By the definition of $\hat{\be}_{i}$,
\begin{eqnarray*}
|\D_{i}^{-1/2}\hat{\S}_{-i,-i}\hat{\be}_{i}-\D_{i}^{-1/2}\hat{\a}|_{\infty}\leq \lambda_{ni}(2).
\end{eqnarray*}
Then it follows that
\begin{eqnarray*}
|\D_{i}^{-1/2}\hat{\S}_{-i,-i}(\hat{\be}_{i}-\be_{i})|_{\infty}\leq 2\lambda_{ni}(2)
\end{eqnarray*}
with probability tending to one. We next prove the restricted eigenvalue (RE)
 assumption in Bickel, Ritov and Tsybakov (2009), page 1710 holds with $\kappa(s,1)\geq c\lambda_{\min}(\S)^{1/2}$ for some $c>0$. Actually,
 the RE assumption follows from $$\max_{1\leq i\leq p}|\be_{i}|_{0}=o\Big{(}\lambda_{\min}(\S)\sqrt{\frac{n}{\log p}}\Big{)}$$ and the inequality
 \begin{eqnarray}\label{tp}
 \de^{'}\hat{\S}_{-i,-i}\de\geq \lambda_{\min}(\S_{-i,-i})|\de|_{2}^{2}-O_{\pr}\Big{(}\sqrt{\frac{\log p}{n}}\Big{)}|\de|_{1}^{2}
 \end{eqnarray}
 for any $\delta\in \R^{p}$. By the proof of Theorem 7.1 in  Bickel, Ritov and Tsybakov (2009), we obtain that
 \begin{eqnarray}\label{tp1}
\max_{1\leq i\leq p}(\hat{\be}_{i}-\be_{i})^{'}\hat{\S}_{-i,-i}(\hat{\be}_{i}-\be_{i})=O_{\pr}\Big{(}\frac{\max_{1\leq i\leq p}|\be_{i}|_{0}\log p}{\lambda_{\min}(\S)n}\Big{)}
 \end{eqnarray}
 and
 \begin{eqnarray}\label{tp2}
 \max_{1\leq i\leq p}|\hat{\be}_{i}-\be_{i}|_{1}=O_{\pr}\Big{(}\max_{1\leq i\leq p}|\be_{i}|_{0}\lambda_{\min}(\S)^{-1}\sqrt{\frac{\log p}{n}}\Big{)}.
 \end{eqnarray}
 This implies Proposition \ref{pro2}.
\qed

\noindent{\bf Proof of Proposition \ref{pro3-3}.}
By the proof of Proposition \ref{pro2}, we have for any $\delta>2$ and some $1<c<\delta/2$,
\begin{eqnarray}\label{tsd}
\max_{i\neq j}\frac{1}{\sqrt{n\hat{\sigma}_{jj}}}\Big{|}\sum_{k=1}^{n}(X_{kj}-\bar{X}_{j})\varepsilon_{ki}\Big{|}\leq \frac{1}{c}\lambda_{ni1}(\delta)
\end{eqnarray}
with probability tending to one.  For a vector $\a=(a_{1},\ldots,a_{p})^{'}$ and an index set $T\subseteq \{1,2,\ldots,p\}$, let $\a_{T}$
be the vector with $(\a_{T})_{i}=a_{i}$ for $i\in T$ and $(\a_{T})_{i}=0$ for $i\in T^{c}$. Let $T_{i}$ be the support of $\be_{i}$.  Then
by the proof of Theorem 1 in Belloni, Chernozhukov and Wang (2011), we can get $|(\hat{\al}_{i}(\delta)-\D_{i}^{1/2}\be_{i})_{T^{c}_{i}}|_{1}\leq \bar{c}|(\hat{\al}_{i}(\delta)-\D_{i}^{1/2}\be_{i})_{T_{i}}|_{1}$ for $\bar{c}=(c+1)/(c-1)$. Also
\begin{eqnarray*}
|\D_{i}^{-1/2}\hat{\S}_{-i,-i}\D_{i}^{-1/2}(\hat{\al}_{i}-\D_{i}^{1/2}\be_{i})|_{\infty}\leq 2\lambda_{ni}(\delta)
\end{eqnarray*}
with probability tending to one. By the proof of Theorem 7.1 in  Bickel, Ritov and Tsybakov (2009), we can get (\ref{tp1}) and (\ref{tp2})
hold for $\hat{\be}_{i}=\hat{\be}_{i}(\delta)$.\qed

\end{document}